\documentclass[sigconf]{acmart}

\AtBeginDocument{%
  }

\setcopyright{acmlicensed}
\copyrightyear{2018}
\acmYear{2018}
\acmDOI{XXXXXXX.XXXXXXX}

\acmConference[Conference acronym 'XX]{Make sure to enter the correct
  conference title from your rights confirmation emai}{June 03--05,
  2018}{Woodstock, NY}
\acmISBN{978-1-4503-XXXX-X/18/06}

\usepackage{tikz}
\usepackage{environ}
\usepackage{soul} 
\usepackage{enumitem}
\usepackage{inputenc}
\usepackage{balance}
\usepackage{fancybox}
\usepackage{framed}

\usetikzlibrary{shapes.misc, positioning}

\tikzset{rightstart fill/.style={append after command={
   \pgfextra
        \draw[sharp corners, fill=bubblegray, bubblegray, text=black, anchor=north west]%
    (\tikzlastnode.west)%
    [rounded corners=6pt] |- (\tikzlastnode.north)%
    [rounded corners=4pt] -| (\tikzlastnode.east)%
    [rounded corners=0pt] |- (\tikzlastnode.south)%
    [rounded corners=6pt] -| (\tikzlastnode.west);
   \endpgfextra}}}

\tikzset{rightend fill/.style={append after command={
   \pgfextra
        \draw[sharp corners, bubblegray, fill=bubblegray, text=black, anchor=north west]%
    (\tikzlastnode.west)%
    [rounded corners=6pt] |- (\tikzlastnode.north)%
    [rounded corners=0pt] -| (\tikzlastnode.east)%
    [rounded corners=4pt] |- (\tikzlastnode.south)%
    [rounded corners=6pt] -| (\tikzlastnode.west);
   \endpgfextra}}}

\tikzset{rightmiddle fill/.style={append after command={
   \pgfextra
        \draw[sharp corners, bubblegray, fill=bubblegray, text=black, anchor=north west]%
    (\tikzlastnode.west)%
    [rounded corners=6pt] |- (\tikzlastnode.north)%
    [rounded corners=0pt] -| (\tikzlastnode.east)%
    [rounded corners=0pt] |- (\tikzlastnode.south)%
    [rounded corners=6pt] -| (\tikzlastnode.west);
   \endpgfextra}}}

\tikzset{rightsingle fill/.style={append after command={
   \pgfextra
        \draw[sharp corners, bubblegray, fill=bubblegray, text=black, anchor=north west]%
    (\tikzlastnode.west)%
    [rounded corners=5pt] |- (\tikzlastnode.north)%
    [rounded corners=5pt] -| (\tikzlastnode.east)%
    [rounded corners=5pt] |- (\tikzlastnode.south)%
    [rounded corners=5pt] -| (\tikzlastnode.west);
   \endpgfextra}}}

\tikzset{leftsingle fill/.style={append after command={
   \pgfextra
        \draw[sharp corners, bubbletan, fill=bubbletan, text=black, anchor=north west]%
    (\tikzlastnode.west)%
    [rounded corners=5pt] |- (\tikzlastnode.north)%
    [rounded corners=5pt] -| (\tikzlastnode.east)%
    [rounded corners=5pt] |- (\tikzlastnode.south)%
    [rounded corners=5pt] -| (\tikzlastnode.west);
   \endpgfextra}}}

\def\mybox#1{\leavevmode \setbox0=\hbox{#1}%
   \dimen0=\wd0 \edef\posxA{\expandafter\ignorept\the\dimen0 \space}%
   \hbox{\kern5pt\pdfliteral{q .8 .8 1 rg .8 .8 1 RG .9963 0 0 .9963 0 0 cm 1 j 1 J 6 w
                             0 0 m 0 5 l \posxA 5 l \posxA 0 l 0 0 l B Q}%
         \box0 \kern5pt}%
}
{\lccode`\?=`\p \lccode`\!=`\t  \lowercase{\gdef\ignorept#1?!{#1}}}

\definecolor{bubblegray}{RGB}{241,240,240}
\definecolor{bubbletan}{RGB}{232, 228, 216}
\definecolor{table-orange}{HTML}{e8a76c}
\definecolor{table-green}{HTML}{a6ec99}
\definecolor{table-blue}{HTML}{97ccf6}
\definecolor{table-purple}{HTML}{d4c0d5} 
\definecolor{table-gray}{HTML}{d3d3d3}
\definecolor{arsenic}{HTML}{36454F}

\newcommand{\readability}[0]{\colorbox{table-blue}{Readability}}
\newcommand{\integrity}[0]{\colorbox{table-green}{Integrity \& Transparency}}
\newcommand{\design}[0]{\colorbox{table-orange}{Quality \& Design}}
\newcommand{\familiarity}[0]{\colorbox{table-purple}{Familiarity}}
\newcommand{\personal}[0]{\colorbox{table-gray}{Personal}}

\newcommand{\clarity}[0]{\colorbox{table-blue}{Clarity}}
\newcommand{\simple}[0]{\colorbox{table-blue}{Simple}}

\newcommand{\dataintegrity}[0]{\colorbox{table-green}{Data Integrity}}
\newcommand{\intent}[0]{\colorbox{table-green}{Intent}}
\newcommand{\source}[0]{\colorbox{table-green}{Source}}

\newcommand{\objectivity}[0]{\colorbox{table-green}{Objectivity}}

\newcommand{\vistype}[0]{\colorbox{table-orange}{Vis Type}}
\newcommand{\visual}[0]{\colorbox{table-orange}{Visual Elements}}
\newcommand{\aesthetic}[0]{\colorbox{table-orange}{Aesthetics}}

\newcommand{\topicfamiliarity}[0]{\colorbox{table-purple}{Familiarity with Topic}}
\newcommand{\sourcefamiliarity}[0]{\colorbox{table-purple}{Familiarity with Source}}

\newcommand{\confidence}[0]{\colorbox{table-gray}{Confidence}}
\newcommand{\reliability}[0]{\colorbox{table-gray}{Reliability}}
\newcommand{\intuition}[0]{\colorbox{table-gray}{Intuition}}

\newcommand\hblue[1]{\colorbox{table-blue}{#1}}
\newcommand\hgreen[1]{\colorbox{table-green}{#1}}
\newcommand\horange[1]{\colorbox{table-orange}{#1}}
\newcommand\hpurple[1]{\colorbox{table-purple}{#1}}
\newcommand\hgray[1]{\colorbox{table-gray}{#1}}

\begin{document}

\title{Trustworthy by Design: The Viewer’s Perspective on \\Trust in Data Visualization}

 \author{Oen McKinley}
 \email{m.oen@wustl.edu}
 \orcid{0000-0002-0931-3068}
 \affiliation{%
   \institution{Washington University in St. Louis}
   \city{St. Louis}
   \state{Missouri}
   \country{USA}
 }

 \author{Saugat Pandey}
 \email{p.saugat@wustl.edu}
 \orcid{0000-0002-7429-6575}
 \affiliation{%
   \institution{Washington University in St. Louis}
   \city{St. Louis}
   \state{Missouri}
   \country{USA}
 }

 \author{Alvitta Ottley}
 \email{alvitta@wustl.edu}
 \orcid{0000-0002-9485-276X}
 \affiliation{%
   \institution{Washington University in St. Louis}
   \city{St. Louis}
   \state{Missouri}
   \country{USA}
 }








\renewcommand{\shortauthors}{McKinley et al.}



\begin{abstract}
    Despite the importance of viewers' trust in data visualization, there is a lack of research on the viewers’ own perspective on their trust. In addition, much of the research on trust remains relatively theoretical and inaccessible for designers. This work aims to address this gap by conducting a qualitative study to explore how viewers perceive different data visualizations and how their perceptions impact their trust.
    Three dominant themes emerged from the data. First, users appeared to be consistent, listing similar rationale for their trust across different stimuli. Second, there were diverse opinions about what factors were most important to trust perception and about why the factors matter. Third, despite this disagreement, there were important trends to the factors that users reported as impactful. Finally, we leverage these themes to give specific and actionable guidelines for visualization designers to make more trustworthy visualizations.
\end{abstract}



\begin{teaserfigure}
    \includegraphics[width=\linewidth]{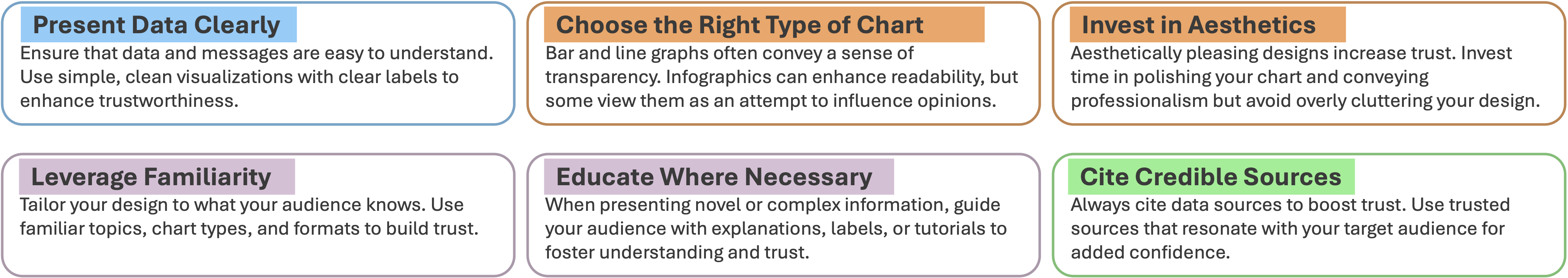}
    \caption{The suggested principles for developing designs that build trust by emphasizing clarity, familiarity, professionalism, and transparency.}
    \Description{A collection of six guidelines for designers to create more trustworthy visualizations. The guidelines include "Present Data Clearly", "Be Mindful of Infographics", "Polish Your Design", "Leverage Familiarity", "Educate Where Necessary", and "Cite Credible Sources".}
    \label{fig:teaser}
\end{teaserfigure}

\received{20 February 2007}
\received[revised]{12 March 2009}
\received[accepted]{5 June 2009}

\maketitle

\section{Introduction}
\label{sec:intro}

Data visualizations are being used more than ever, from business to public health~\cite{leung2020covid, khanam2020covid, shaikh2023covid, islam2019vis, padilla2022multiple, romano2020scale}. As data visualization has expanded, the importance of quality design has become ever more pronounced. Especially in high-risk situations like COVID-19, being able to communicate data effectively is vital to spreading much-needed information~\cite{leung2020covid, khanam2020covid, padilla2022multiple, romano2020scale}. Academic researchers and industry professionals have researched ways to make visualizations more appealing or effective~\cite{tufte20012e, kelleher2011guidelines, midway2020principles, evergreen2013principles}. Recently, there has also been a push that recognizes \textit{trustworthiness} as an essential component of visualization design~\cite{artz2007survey, chita2021can, mayr2019trust}. Eliciting trust, according to these works, is important for any visualization designer. 

Several studies have sought to isolate the factors that influence trust in visualizations. For instance, Elhamdadi et al.~\cite{elhamdadi2023vistrust} introduced a multidimensional framework that categorized trust antecedents into cognitive and affective dimensions and further distinguished between trust in the visualization itself and trust in the underlying data. Pandey et al.~\cite{pandey2023you} took a design-oriented approach, proposing five factors likely to influence trust: credibility, clarity, reliability, familiarity, and confidence. Crouser et al.~\cite{crouser2024building} explored the relationships between endogenous factors (such as visual metaphors, color usage, and source) and exogenous factors (including individual differences in personality, cognitive ability, educational background, and cultural influences) and how these combined to affect trust perception. 

Taken together, this body of prior work offers two major takeaways. First, users may often associate aesthetic quality with professionalism and trustworthiness, even when embellishments do not enhance clarity or accuracy~\cite{pandey2023you,crouser2024building}. Therefore, designers need to understand how to improve the trustworthiness of their visualizations with embellishments and avoid the risk of obscuring or distorting underlying data. 
Second, personal factors may influence perceived trust~\cite{elhamdadi2023vistrust,crouser2024building}. Users come from diverse backgrounds with varying expectations regarding how information should be presented~\cite{liu2020survey,ottley2022adaptive}, and these individual differences may further mediate how users assess trustworthiness, making it difficult to develop a one-size-fits-all approach~\cite{kim2017data, kong2019misalign, zehrung2021vis, zhou2019effects,elhamdadi2022using,crouser2024building}. 

There is clearly increasing recognition of the importance of trust in visualization design, and much of the literature implicitly discusses how to make a visualization more trustworthy. However, there remains a notable gap in developing clear, consistent design principles related to trust that can be broadly accessible to visualization designers and industry professionals. While guidelines for data visualization exist, primarily emphasizing elements such as selecting appropriate visual encodings for a given dataset or task and aesthetics~\cite{kelleher2011guidelines, midway2020principles, evergreen2013principles, tufte20012e}, \textit{there is no design framework tailored explicitly to increase the trustworthiness of visualizations across different user groups}.
This paper seeks to address these challenges by identifying consistent principles that can guide designers in creating visualizations that are broadly perceived as trustworthy. We synthesize insights from an exploratory qualitative study on trust perception with three primary goals: determine whether there are common factors that contribute to trustworthiness, regardless of user background; explore how individual differences, especially regarding education and design experience, influence users' trust perceptions; and provide specific guidelines that designers can apply to enhance the trustworthiness of their visualizations.





Our analysis uncovered the following:

\begin{itemize}
    \item \textbf{Internally Consistent Trust Perception:} When analyzing responses at the individual participant level, we find evidence that each of our participants had their own internally consistent frameworks for rationalizing their trust perception. This means that each individual will often consistently look for the same factors when determining trust and will often consider the same visualization in the same way through different contexts. 

    \item \textbf{Diverse Internal Frameworks:} When analyzing responses between participants, we find evidence that different participants describe a diverse range of internal frameworks. In other words, participants often disagree about what factors are important in their decision, and even when their frameworks overlap, the metrics they use to evaluate those factors may not align or even be contradictory.

    \item \textbf{Aggregated Trends in Trust Perception:} Despite disagreement between specific individuals, when analyzing responses at a comprehensive level, there are still trends in the frameworks described by participants. These trends, while not universal, offer important insight into the areas where visualizations can gain trust from a broad range of users.
\end{itemize}

Finally, we contribute a set of specific guidelines based on the results of this work. These guidelines are explicitly intended to help designers create \textit{trustworthy} visualization designs.

\section{Background \& Related Work}
\label{sec:background}

This work builds on two important branches of work in data visualizations: Design Guidelines and Trust. Although both of these branches contain a wealth of information relevant to our goals, their intersection offers an opportunity to apply the research in data visualization trust to a more digestible format that visualization designers can use. 

\subsection{Visualization Design Guidelines}

Since the popularization of digital data visualization, there have been efforts to standardize theoretically derived or empirically tested design recommendations into specific, useful guidelines for designers~\cite{tufte20012e, kelleher2011guidelines, midway2020principles, evergreen2013principles, franconeri2021works}. 

Edward R. Tufte is arguably one of the most influential figures in this history. In his book \textit{The Visual Display of Quantitative Information}~\cite{tufte20012e}, he popularized many now-common visualization guidelines such as ``above all else, show the data'' and the data-ink ratio. While many of Tufte's guidelines have been updated in recent years to reflect modern trends or new empirical evidence, his influence on the field is undeniable. 
More recent works include Kelleher \& Wagener~\cite{kelleher2011guidelines} and Evergreen \& Metzner~\cite{evergreen2013principles}, both of which build on prior work to provide guidelines and categorize them into themes such as simplification and emphasis. 
This has also led more authors, such as Midway~\cite{midway2020principles}, to provide even more up-to-date lists of various guidelines and principles for designing higher-quality visualizations. 
Similarly, there are also more specific use cases for visualization guidelines. For example, Lin et al. \cite{lin2023hunches} provide guidelines for visualizing ``data hunches'' to help experts communicate insights into the data that may otherwise be difficult to express.
Furthermore, recent research has also pushed for a stronger focus on empirically tested visualization design principles.
Franconeri et al.~\cite{franconeri2021works}, in particular, provide a review of empirical studies showing the perceptual accuracy of different visual channels (such as position or length), common illusions in various charts, cognitive biases, and color-vision impairments, among others.

Such work not only furthers research into what design features make visualizations more appealing or effective but also provides resources to designers who may not have a background in the depths of visualization literature. However, there is a lack of research on either formulating new guidelines specifically tailored toward audience trust perception or applying existing guidelines to the trustworthiness of a given visualization. Applying design guidelines would be an important step in furthering the communication of trust-related concepts to a broader community of designers.

\subsection{Defining Trust in Data Visualization}

Researchers have long sought to understand how individual differences and personality traits shape performance, highlighting the importance of tailoring visualization design to diverse audiences~\cite{liu2020survey, ottley2022adaptive}. 
Trust, in the context of personality, refers to an individual’s expectation that others’ statements—whether spoken, written, or visual—can be relied upon~\cite{rotter1967new}. Work by Peck et al. suggests that this trait can play a critical role in determining whether audiences perceive data visualizations as trustworthy. Using in-depth interviews in rural Pennsylvania, they gathered insights on how personal traits and life experiences shape trust in data visualizations~\cite{peck2019data}.

Mayr et al. sought to define trust in visualization by separating it into two components: \textit{trustworthiness} and \textit{trust perception}~\cite{mayr2019trust}. They define trustworthiness as ``the properties of the visualization (and the underlying data) that lead users to trust it.'' Conversely, trust perception, ``is the user’s subjective evaluation of the quality and reliability of the visualized information''.
In other words, the trustworthiness of visualization and the trust perception of a user are two heavily related yet distinct topics. The end goal of a visualization designer is to gain the user's trust. However, they cannot directly manipulate the users' trust perception and must, therefore, adjust the trustworthiness of the visualization as a proxy. Therefore, it is imperative for the designer to know how the trustworthiness of a visualization maps to its design elements.

\subsection{Prior Work on Trust in Data Visualization}

Researchers have long acknowledged the complexities of visualization trust and its significance, emphasizing its role in data interpretation and effective communication~\cite{artz2007survey, chita2021can, mayr2019trust}. However, works often note the complexities when defining and measuring trust in data visualization~\cite{mayr2019trust, sacha2015role, xiong2019examining, pandey2023you, elhamdadi2023vistrust, mcallister1995affect}. 

Mayr et al.~\cite{mayr2019trust} were among the first to advocate for an atomistic approach to gauging trust in data visualization. Pandey et al.~\cite{pandey2023you} furthered this work by adopting a design-centric approach to identify five key factors influencing trust in visualizations: credibility, clarity, reliability, familiarity, and confidence.
Their findings revealed that participants rated colorful visualizations with visual embellishments more favorably regarding the five trust factors. Interestingly, visualizations from news media were perceived as more credible and reliable than those from scientific or governmental agencies, even when explicit information about the source was not provided. This suggests that the preference towards news media visualizations was not simply a bias against government or scientific institutions but rather a preference for the design used by news agencies over the designs used by other sources.
This work was then built upon by Crouser et al.~\cite{crouser2024building}, who separated the factors leading to trustworthiness into endogenous factors (such as visual metaphors, color usage, and source) and exogenous factors (including individual differences in personality, cognitive ability, educational background, and cultural influences). Their findings provided evidence that visualization design has a significant relationship with users' trust perception, and visualization literacy was the strongest predictor of deviations in trust. 
Elhamdadi et al.~\cite{elhamdadi2023vistrust} also independently identified and addressed this gap by proposing a multidimensional framework, building on the work of McAllister~\cite{mcallister1995affect} and characterizing trust as either \textit{affective-based} (related to aesthetics and ethical alignment) or \textit{cognition-based} (related to clarity, accuracy, and usability). Additionally, they distinguish trust in the data from trust in the visualization design. While these efforts have contributed invaluable perspectives on trust, they also have a common theme of focusing on theoretical constructs of how trust should be defined.

One obvious way to build trust is to avoid design choices that could be seen as misleading, such as truncating axes \cite{lo2022misinfo, pandey2015deceptive, lisnic2023lie}, omitting or selectively representing data \cite{lo2022misinfo, lisnic2023lie, kong2019misalign, mcnutt2020mirage}, mismatched titles or text \cite{kong2019misalign, lisnic2023lie}, or many other ``misleaders'' \cite{lo2022misinfo, mcnutt2020mirage, lisnic2023lie}.
However, there has been research on how data visualization affects trust in specific institutions or contexts.
For example, Padilla et al. \cite{padilla2022multiple} investigate the relationship between the use of different data visualizations and an audience's trust in COVID-19 prediction forecasts, finding that audiences were more likely to trust the predictions when simpler visualizations and certain encodings were used. Yang et al. also examined the role that different visualizations have in eliciting trust in the context of the United States 2022 midterm elections \cite{yang2023midterms} and explored the ways that different uncertainty displays elicit trust in United States election scenarios \cite{yang2024dice}. While these works provide valuable insights into trust-building, identifying specific design features that may foster trust in the visualizations themselves was not their primary goal.

However, all of these prior works crucially leave a gap for the practical perspective that users have towards design factors of visualization and the impact on their own trust perception. 
We believe that this practical user perspective is as important to any definition or deconstruction of trust as a theoretical framework and is, therefore, vital to the efforts to analyze trust in the data visualization field.

\section{Methodology}
\label{sec:method}

Previous research suggests that the design of visualizations can significantly influence its perceived trustworthiness~\cite{crouser2024building,pandey2023you}. Therefore, understanding the factors that shape trust is essential for creating visualizations that effectively convey information. We conducted a qualitative user study to identify and understand the design elements that users report as affecting their trust perception.

Trust, however, is not always a straightforward decision. Cognitive science research has extensively explored human reasoning and decision-making, and one influential framework, \textit{dual-process theory}~\cite{kahneman_book,evans2013dual}, posits that human thinking operates through two distinct systems: \textit{Type 1} processes, which are fast, intuitive, and automatic, and \textit{Type 2} processes, which are slower, more deliberate, and analytical. In the context of visualizations, both processes may influence decisions, including trust~\cite{bancilhon2023eval}. This paper focuses on Type 2 trust — the more intentional, reflective decision-making process.
We hypothesize that when participants engage in deliberate trust, they are more likely to reflect on the specific features or qualities of a visualization that shape their decision to trust it.
While subconscious trust remains important, it is beyond the scope of this work. Our approach, therefore, should explore explicit insights into which design elements contribute to the users' trust perception to provide specific and useful feedback for more trustworthy visualizations.

\subsection{Survey Design}
\label{sub:survey}

This exploratory study aimed to uncover what design features make visualizations appear trustworthy to viewers. To achieve this, we employed a mixed-methods approach that combined quantitative ranking tasks with qualitative analysis of participant explanations. We recruited 40 online participants to ensure a broad geographic representation and a range of perspectives, improving the generalizability of our findings.

Participants were shown visualizations from four different sources: news outlets, scientific venues, government publications, and infographics. This diverse selection was intentional to reflect the various contexts in which people encounter visualizations in daily life. 
Each participant was presented with five groups of six visualization designs, some of which were repeated across rounds to assess consistency in rankings. The ranking provided participants with a frame of reference for evaluating trustworthiness while minimizing the risk of overly influencing their opinions. Within each group, participants were asked to rank the visualizations based on perceived trustworthiness. After completing the ranking tasks, participants were asked to explain the reasoning behind their rankings. This open-ended question provided rich qualitative data, revealing the specific design features and contextual factors that influenced their trustworthiness judgments. We analyzed the rankings quantitatively to identify patterns in trustworthiness perceptions across design categories and participant groups.  The qualitative data from participant explanations were thematically analyzed to uncover which design features were frequently mentioned and how they contributed to trustworthiness perceptions. 

This study design balances the need for contextual grounding with the goal of minimizing bias. By providing participants with concrete visualization examples, we ensured their responses were anchored in real-world designs without directly steering their judgments. The inclusion of diverse sources broadened the applicability of the findings, while the ranking task offered a manageable yet insightful way to capture comparative judgments.
Alternative designs, such as free-form discussions or a focus on a single design source, were rejected because they either lacked structure for systematic analysis or risked limiting the scope of insights. Similarly, a purely quantitative approach would have missed the depth of understanding provided by participants’ explanations.
Overall, this mixed-methods approach allowed us to capture both the subjective judgments of trustworthiness and the objective design features influencing those judgments, offering valuable insights into the interplay between visualization design and audience trust.

\subsection{Visualization Selection}
\label{sub:choicevis}

Our choice of visualizations was informed by the work of Pandey et. al~\cite{pandey2023you}, who identified five key dimensions of trust in data visualizations: \textit{Clarity, Credibility, Familiarity, Reliability, and Confidence}. The visualizations were taken from the MASSVIS dataset \cite{borkin2013makes}, which were collected from a diverse set of sources such as government agencies, news venues, scientific publications, and infographics.
By adopting their visualization set, we ensured that our stimuli varied across these important dimensions. For example, a single round might include familiar and less common chart types, visualizations with varying levels of visual clarity, and representations of data with different levels of reported credibility or reliability. This approach allows us to observe how participants weigh these different factors when making trust judgments with designs that represent a range of contexts in which people encounter visualizations in their daily lives. The specific visualizations used in our study are illustrated in \autoref{fig:stimuli} in \autoref{appendix-stimuli}.

\subsection{Participants and Study Procedure}
\label{sub:participants}

\begin{table}[t]
    \centering
    \caption{The demographics of the 40 participants.}
    \begin{tabular}{lccc}
        \hline
        \textbf{Characteristic} & \textbf{Count} & \textbf{Percentage} \\
        \hline
        Gender & & \\
        \hspace{0.3cm} Male & 21 & 52.5\% \\
        \hspace{0.3cm} Female & 18 & 45\% \\
        \hspace{0.3cm} Non-binary/Third gender & 1 & 2.5\% \\
        Education & & \\
        \hspace{0.3cm} Doctorate degree & 1 & 2.5\% \\
        \hspace{0.3cm} Professional degrees & 2 & 5\% \\
        \hspace{0.3cm} Two-year degrees & 6 & 15\% \\
        \hspace{0.3cm} Four-year degree programs & 14 & 35\% \\
        \hspace{0.3cm} Some college education & 11 & 27.5\% \\
        \hspace{0.3cm} High school graduates & 6 & 15\% \\
        Visualization Creation Experience & & \\
        \hspace{0.3cm} Never created visualizations & 12 & 30\% \\
        \hspace{0.3cm} Somewhat familiar & 16 & 40\% \\
        \hspace{0.3cm} Prior experience & 12 & 30\% \\
        Age & & \\
        \hspace{0.3cm} 18--29 & 14 & 35\% \\
        \hspace{0.3cm} 30--39 & 15 & 37.5\% \\
        \hspace{0.3cm} 40--49 & 5 & 12.5\% \\
        \hspace{0.3cm} 50--60 & 6 & 15\% \\
        \hline
    \end{tabular}
    \label{tab:participant-demographics}
\end{table}

We recruited 40 participants through Prolific for the survey. All participants were from the United States, spoke English, and held a Prolific approval rating of at least 95\%. \autoref{tab:participant-demographics} presents a comprehensive overview of the participants' demographic breakdown.

\begin{figure*}[h]
    \centering
    \includegraphics[width=\linewidth,alt={A hierarchy of our codebook. The keywords are separated into five categories: \readability, \integrity, \design, \familiarity, and \personal.}]{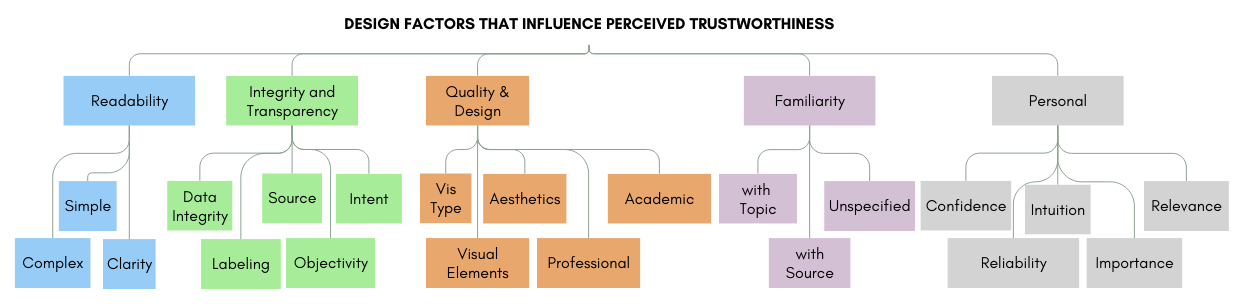}
    \caption{The categorization of keywords in our codebook. For definitions of each keyword, see \autoref{tab:codedefs} in \autoref{appendix_codebook}.}
    \label{fig:key-cats}
\end{figure*}

\label{sub:finaldesign}

After accepting the task through Prolific, participants were redirected to Qualtrics to complete the survey. The survey consisted of three parts: a consent form, five rounds of visualization comparisons, and a demographic questionnaire. The order of the five rounds was randomized, but the visualizations presented in each round were static for ease of analysis across users. The participants would then see the six visualizations in the given round, each with an identifier ``a'' through ``f'' (which were also not randomized), and be asked to rank these designs from 1 to 6, without the ability to give the same rank to two visualizations. This forced the users to rank visuals higher or lower than one another. After completing the ranking exercise, a long-form text box below prompted the users to ``Please explain your ranking.'' This was repeated for all five rounds, such that all visualizations were ranked by all participants.

\subsection{Responses \& Coding}
\label{sub:responses}

To independently categorize the participants' open-ended responses regarding their trust rankings based on similarity and shared attributes, we used a theoretical coding methodology inspired by Grounded Theory \cite{hull2013gt}. Inter-rater reliability (according to a review performed by Bajpai et al.~\cite{bajpai2015rating}) was then used to evaluate the agreement and reconcile the codebooks. First, one author reviewed the participants' responses and conducted a round of open coding to generate an initial taxonomy of trust antecedents. Then, two authors independently reviewed all 200 explanations and performed a second round of open coding, assigning relevant codes where applicable. This independent coding process assessed the initial consistency and agreement in code assignment, producing an inter-rater reliability of 0.44. This score indicated moderate agreement between the two coders but highlighted discrepancies to be addressed. The two authors then collaborated to review and discuss the coding assignments to resolve these differences. Through this collaborative effort, they reached a consensus on code assignments, leading to the refinement of the codebook. For a list of all keywords and definitions used, refer to \autoref{tab:codedefs} in \autoref{appendix_codebook}. The keywords were then sorted into categories with core similarities for discussion in this paper (\autoref{fig:key-cats}).

After the coding stage of the analysis, 3 participants were removed (Participants 6, 8, and 27) due to a lack of meaningful information in their responses. Therefore, there were 37 participants (185 responses) in the final analysis.

\section{Analysis \& Results}
\label{sec:analysis}

To gain the most potential insight into the participants' trust perception, we analyze three levels of granularity. Each of these levels yields a different perspective on users' trust perception. Specifically, there are three different themes that are visible when looking at these different viewpoints:

\begin{enumerate}
    \item \textbf{\textsc{Individual Level}} \hspace{.75em}By analyzing responses at the individual level, we uncovered that participants are largely consistent in their trust perceptions. 
    \item \textbf{\textsc{Group Level}} \hspace{.75em}When comparing individuals, there were often different (sometimes contradictory) factors that influenced each participant's trust perception.
    \item \textbf{\textsc{Holistic Level}} \hspace{.75em}We examined the responses holistically for broad trends. Despite disagreement individually, there were still several popular takeaways from the responses.
\end{enumerate}

Finally, we also examine the potential role that demographic factors had in the responses given by participants.

\begin{figure*}[h]
\centering
\includegraphics[width=\linewidth, alt={A stacked bar chart displaying the keywords used by each participant in each of the five rounds. Colors correspond to the keyword category, and a recycle symbol shows when keywords have been reused.}]{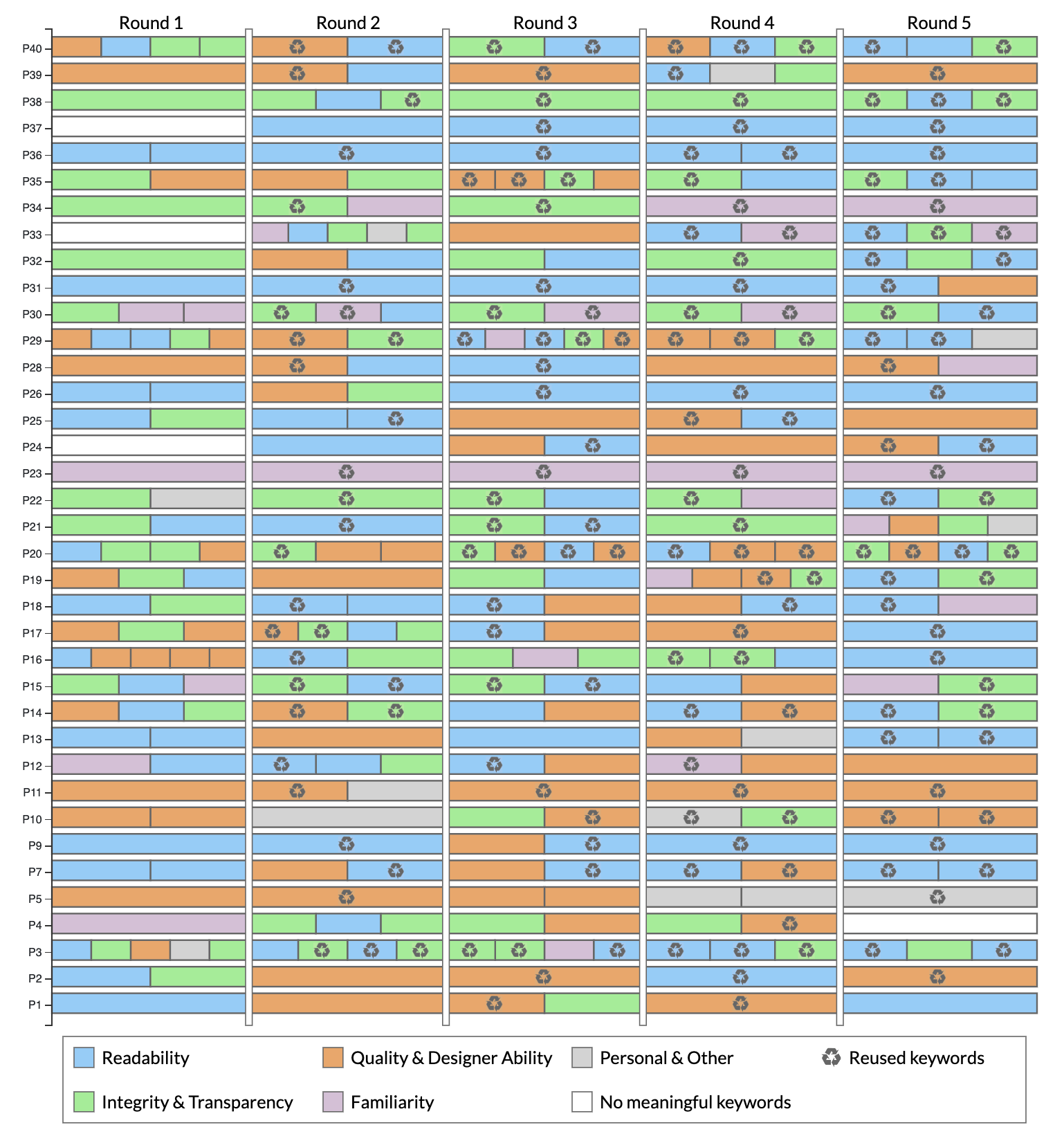} 
\caption{
The keywords used by each participant are colored based on the keyword category. Although the order of the rounds was randomized for each participant, we have aligned them in this figure to illustrate what participants attended when ranking the same charts. Participants used a small set of keywords, though the specific set varied between individuals.\\ 
}
\label{fig:participantfreq}
\end{figure*}

\subsection{Theme 1: Response Consistency within Individual Participants}
\label{sub:theme1}

\subsubsection{Participants attended to a small set of visualization features.} One clear takeaway from participants' responses is internal consistency within each participant. Each participant used a small subset of factors as heuristics to determine trustworthiness. On average, each person used 1.8 keywords in each round to explain their trust perception, meaning an average of 9 total keywords across the five rounds. However, each participant used an average of only 4.6 unique keywords when reasoning about the visualizations across all five rounds. 12 of the 37 participants (32.4\%) had at least one keyword they reused in all five rounds, and 20 of the 37 participants (54.1\%) had at least one keyword they repeated in at least four rounds. 

One prominent example of this pattern is Participant 7 (P7), whose responses consistently emphasized the role of \readability~ in their evaluation process. Across all five rounds of feedback, P7 repeatedly highlighted their ability to understand the visualization as the primary factor influencing their trust in it:







\begin{flushright}
\includegraphics[width=\linewidth]{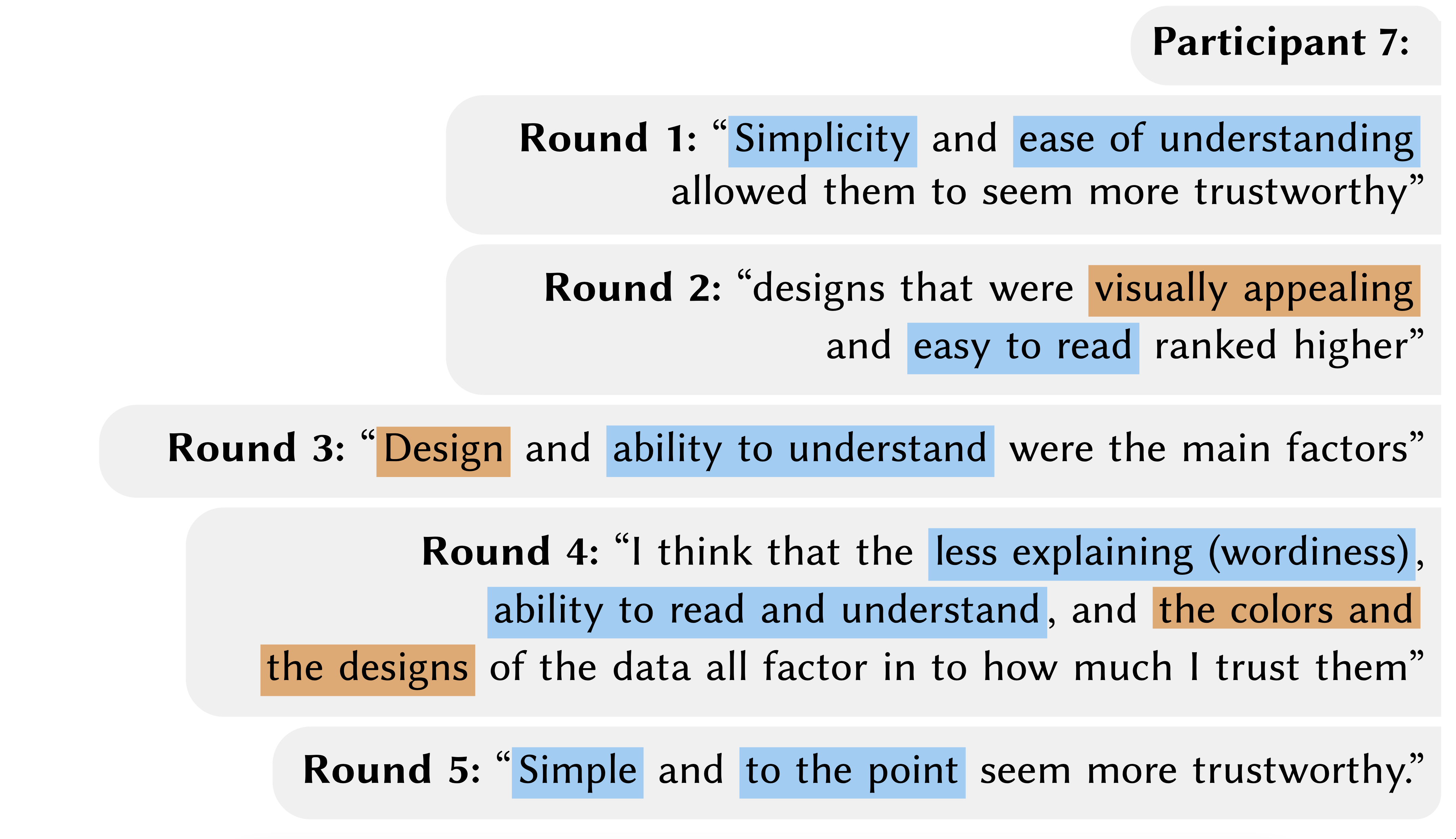}
\end{flushright}


Like many participants, these comments show P7's predisposition to trust simple and readable visualizations.
As in \autoref{fig:key-cats}, we categorize these preferences as \readability, the most common category in our coding process used by 32 of our 37 participants (86.5\%). P7 was also concerned with the \design~ displayed in the visual (including colors and aesthetic appeal), and this consistent emphasis on both \readability~ and \design~ reflects a simplicity and consistency behind their framework for trust perception.

Another example, P15, was more concerned with source citation for the data in the visualizations, which we coded under \integrity. Of the five responses they left, four explicitly referenced the presence of source citations:







\begin{flushright}
\includegraphics[width=\linewidth]{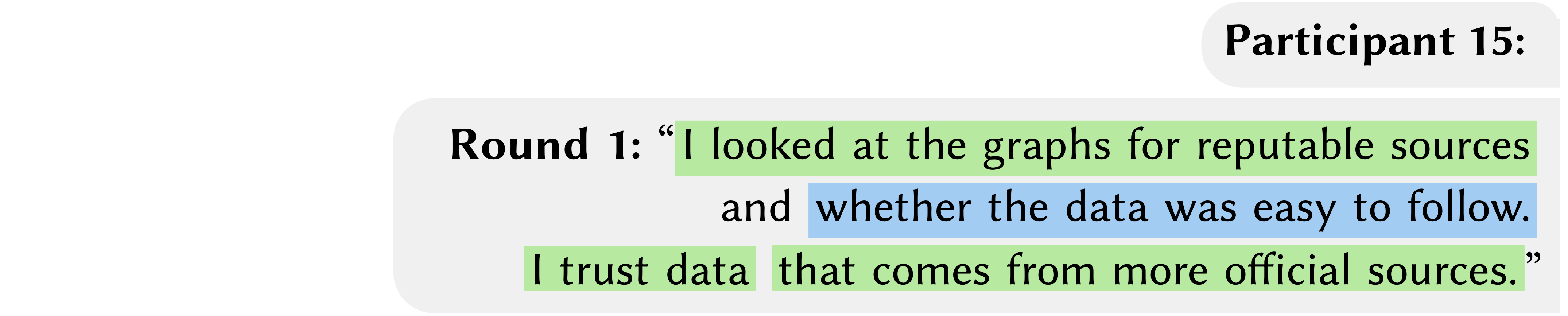}
\end{flushright}

\begin{flushright}
\includegraphics[width=\linewidth]{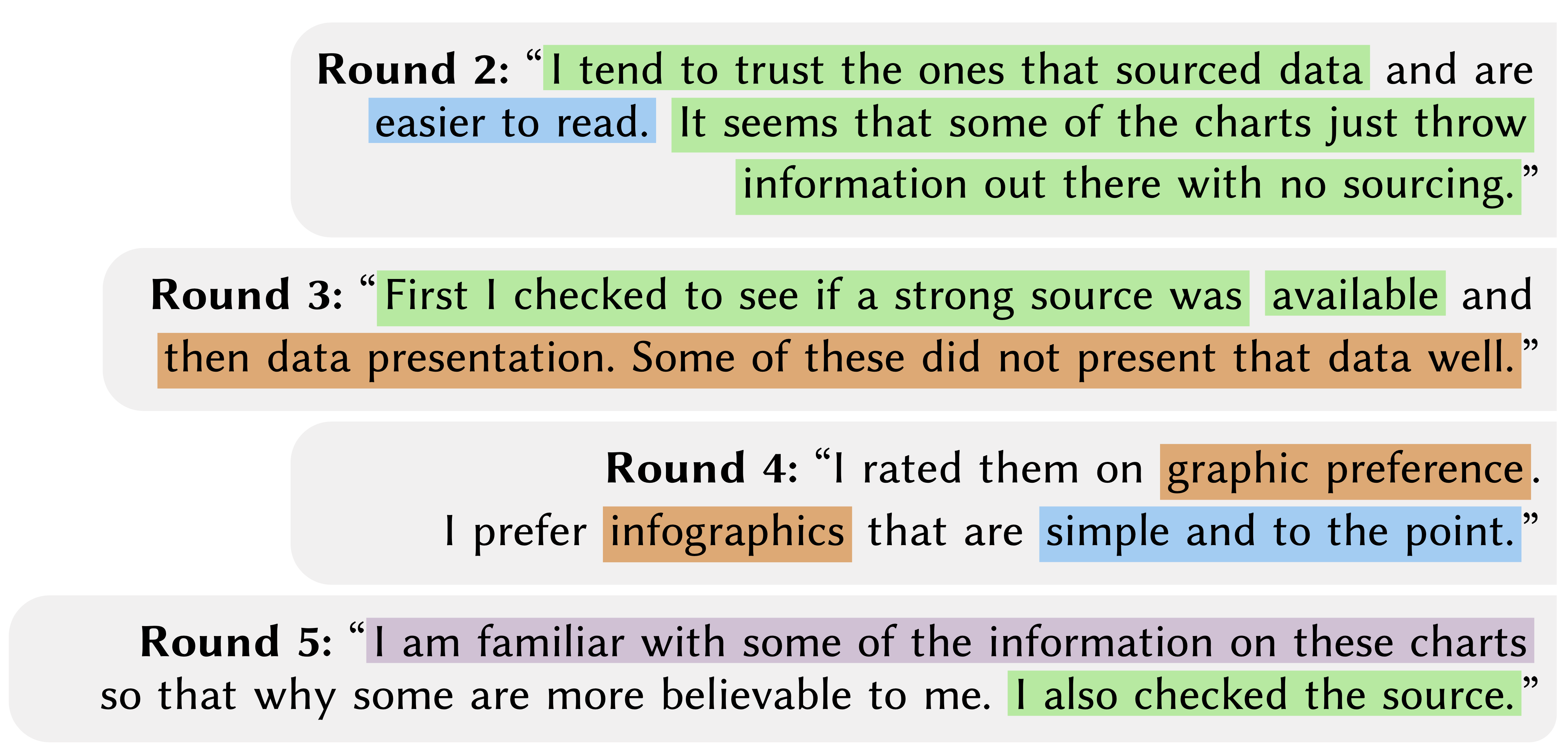}
\end{flushright}

Like several other participants, P15 expressed greater trust in visualizations that explicitly indicated where the data originated and implied that they actively sought out strong, reliable sources. This reliance on sources underscores a broader trend among participants: the perception of trust was closely tied to whether or not the visualization demonstrated transparency in its data origins.

Aside from these examples, most participants also followed a similar trend of keyword reuse: \ul{for the majority of individual participants, a small subset of factors tended to consistently be employed throughout repeated rounds and different contexts.} \autoref{fig:participantfreq} displays the reuse of individual keywords (with even more granularity than general categories). All participants reused keywords at least once, and most participants reused keywords in multiple rounds. At the category level (represented by the colors in the figure), every one of the 37 participants had at least one category of response repeated in at least three rounds, with 15 of the 37 participants (40.5\%) reusing a category in all five rounds.
The observed consistency in most participants' responses, despite different visualizations being discussed, suggests that it is feasible to tailor visualizations to an individual. Designers can leverage this predictability to create visualizations that align with users' expectations, enhancing the user experience. Moreover, since participants relied on a limited set of factors to assess trustworthiness, designers can focus their efforts on optimizing these key factors. 
However, although a narrow focus on elements such as \source\ or \clarity\ can make it easier to tailor designs for a specific individual, it highlights a potential limitation in how participants assess visualizations. By placing their trust in a small subset of factors, participants may unintentionally overlook other critical aspects. Users could benefit from a more holistic approach to evaluating visualization, leading to a more balanced and accurate assessment of visualization trustworthiness. 

\subsubsection{Participants remembered charts and maintained their beliefs} 

It is well-documented that individuals often make different choices when presented with the same options multiple times, a phenomenon known as \textit{stochastic choice}~\cite{tversky1969intransitivity,ballinger1997decisions,alos2021choice}. Although this was not the primary focus of our study, our decision to repeatedly present various visualization designs in different comparison contexts provided a unique opportunity to observe participants' reactions over time. Interestingly, we found that participants' responses were generally stable when they encountered the same visualizations across multiple rounds, offering valuable insights into the consistency of their trust judgments.

P29 not only recognized repeated visualizations but also maintained their initial evaluations throughout the study. Their rankings in both instances reflected a similar reasoning process, suggesting that their trust evaluations were based on consistent criteria:

\begin{flushright}
\includegraphics[width=\linewidth]{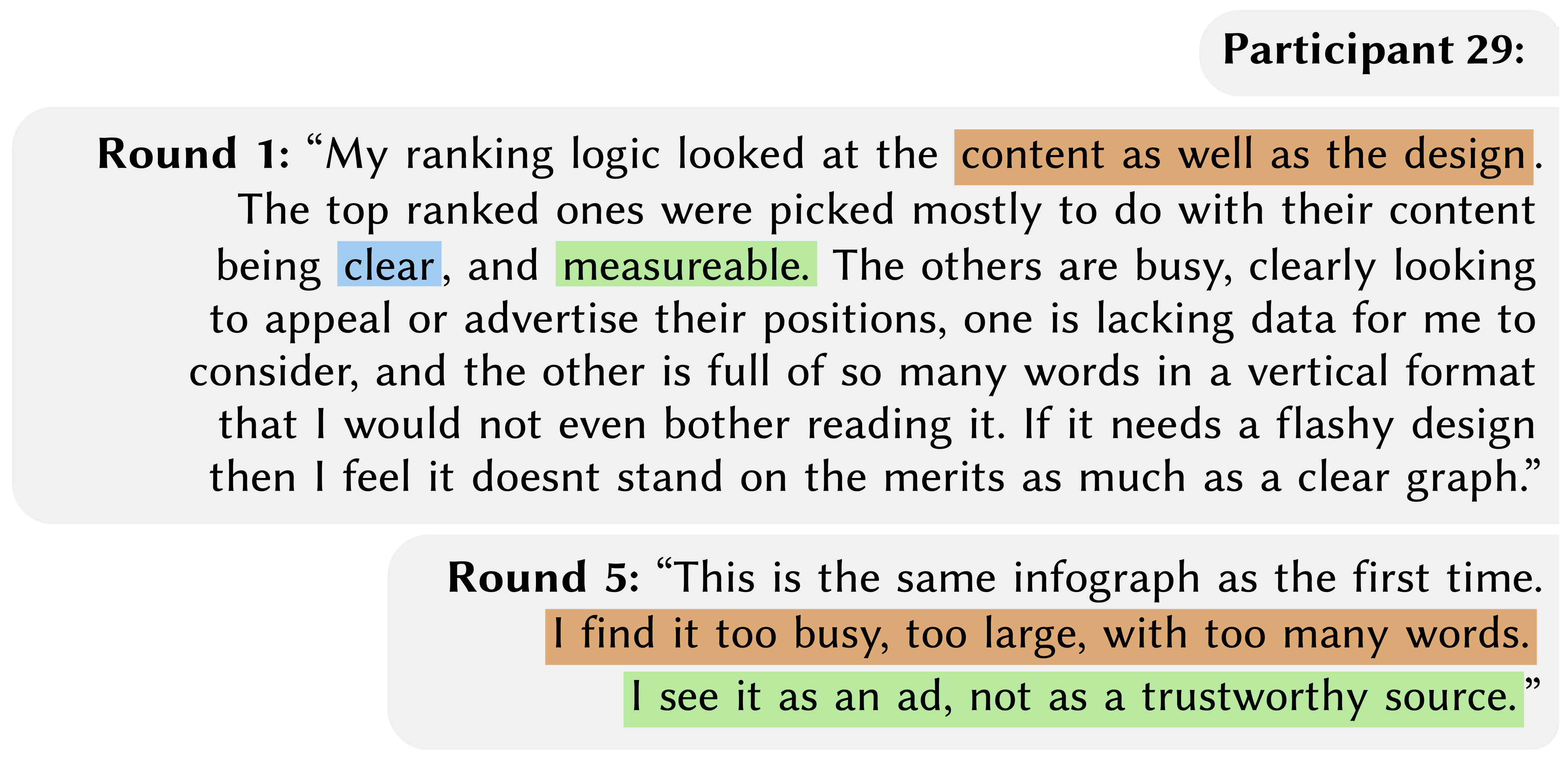}
\end{flushright}

In both of these rounds, P29 focused on the same aspects: content clarity, design simplicity, and the overall layout of the visualization. Their repeated negative assessment of the infographic’s busy design and excessive text, coupled with their preference for clear, data-driven visuals, demonstrates how their trust judgments were influenced by a consistent strong aversion to designs that appeared more promotional than informational.

P3 also exhibited consistency in their evaluations. After initially rating a chart poorly in the first round, P3 recognized the same chart in a later round and reaffirmed their earlier criticism, providing insight into their stable judgment criteria:




\begin{flushright}
\includegraphics[width=\linewidth]{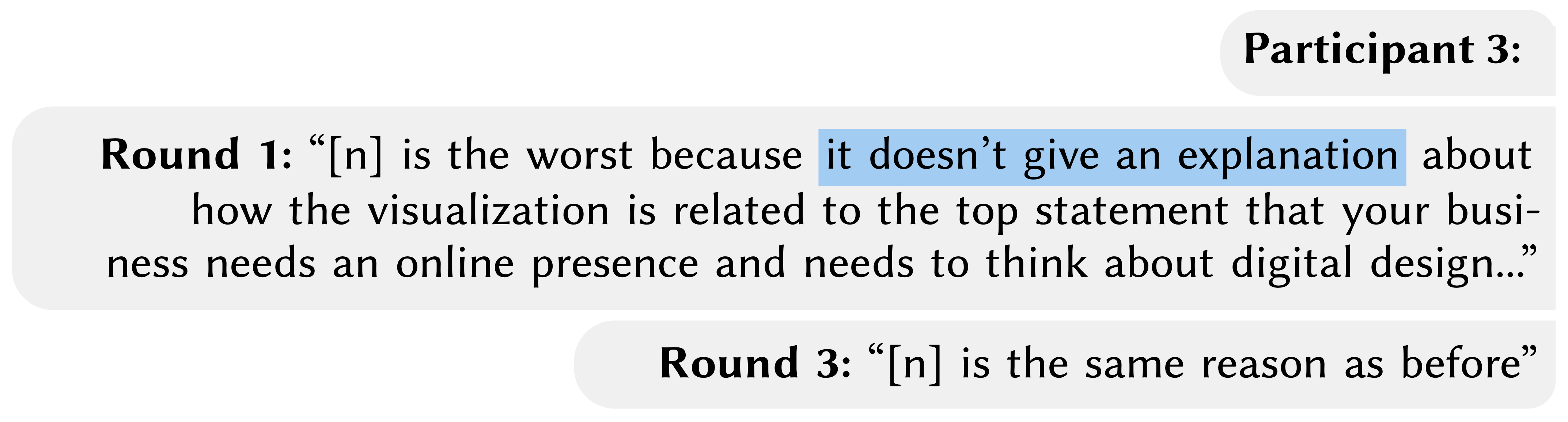}
\end{flushright}

P3’s repeated criticism highlights the lack of explanation and their focus on the relevance of the content to the visual’s stated message. Their consistent dismissal of the visualization for failing to connect the visual representation to the main point demonstrates that trust, for P3, was closely tied to how well the visual fulfilled its communicative purpose. This represents a consistent finding that \ul{several participants both recalled visualizations and maintained consistent trust judgments about said visualizations.}

Interestingly, in both instances, the visualizations in question were infographic designs. This aligns with prior research suggesting that visualizations featuring human-recognizable objects, such as infographics, are more memorable than abstract designs~\cite{borkin2013makes}. It is possible that this enhanced memorability also led participants to recall and evaluate their trustworthiness more consistently over time. Memorability may also be associated with \familiarity, which was proposed as a trust factor by Pandey et al.~\cite{pandey2023you} and was one of the keyword categories revealed in the coding for this study.
However, we reaffirm that memorability was not an explicit focus of our study, so it is unclear whether this consistency was unique to infographics or would extend to other types of visualizations.
Future research could explore whether other visual elements, such as the presence of human-recognizable objects, color schemes, or data complexity that impact memorability, in turn influence trust.

\subsection{Theme 2: Response Inconsistency between Different Participants}
\label{sub:theme2}

\subsubsection{Diverging Priorities.}
While individual participants demonstrated consistency in the factors used when evaluating visualizations, there were notable differences in how these factors were prioritized between individuals. The perceived trustworthiness of a visualization design was subjective and varied widely from one person to another. 
Using \integrity\ as a running example for this section, we can observe from \autoref{fig:keywordfreq} that 26 of the 37 participants (70.3\%) highlighted its importance in assessing trustworthiness. However, their interpretations of these concepts differed; participants did not always focus on the same elements. Some were more concerned with the integrity of the data itself and whether the information seemed accurate and reliable. P3, for example, preferred visualizations that clearly disclosed the data source. Others, like P14, were also concerned with \integrity, but instead prioritized the intent of the designer, feeling wary of designers ``hiding data'' from viewers:





\begin{flushright}
\includegraphics[width=\linewidth]{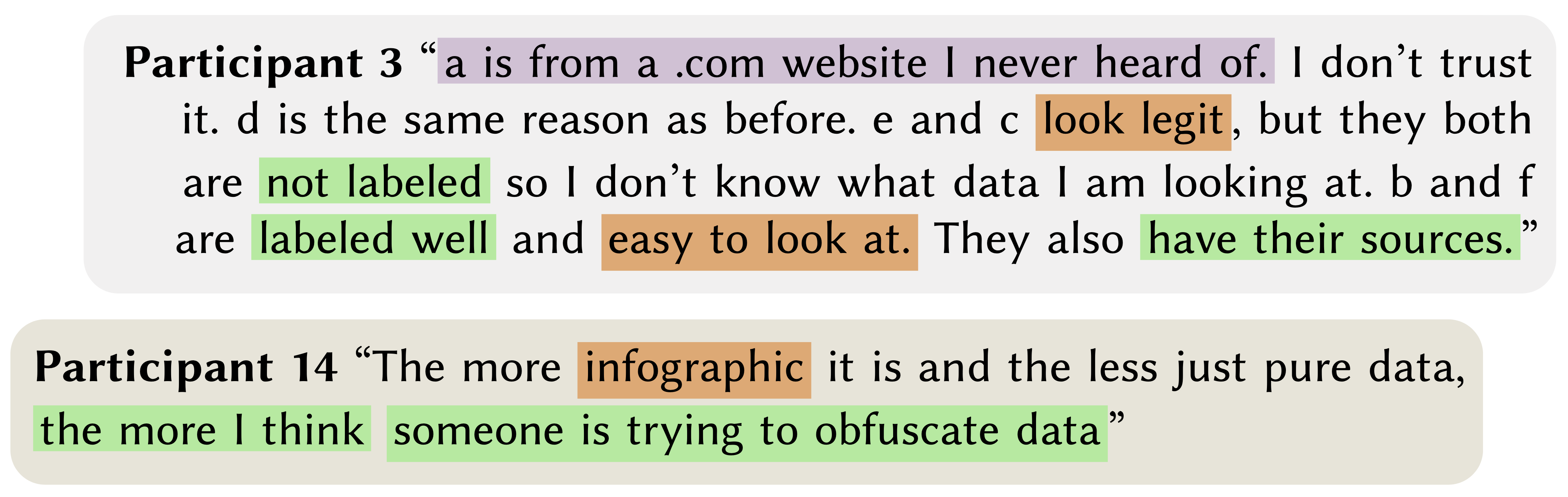}
\end{flushright}

Both of these participants agree that \integrity~ is a critical factor, but their focus on what constitutes transparency differs significantly. One participant emphasizes the importance of clear, cited sources to trust the validity and accuracy while the other participant focuses on the visual encoding of the data, believing that ``pure data'' representations are key to transparency. P14 never mentioned the data source, P3 never questioned whether a visualization guided viewers toward a particular conclusion or agenda. 

The difference in these participants’ focus highlights that, while transparency is a shared priority, what transparency means to each varies.
\ul{It's not enough to simply include a credible source or ensure a visually appealing design; for some users, the way the data is visually encoded and the neutrality of its presentation is just as important in fostering trust.}

Although we use the \integrity\ category to show the divergence of viewers' perspectives, we observed a similar pattern with the other categories. 
\ul{Though many participants agree on the importance of particular concepts, each participant uses different heuristics for evaluation.} This further reinforces that there may be no one-size-fits-all approach to trustworthy design.

\subsubsection{Infographics were contentious.}
In addition to differing priorities or strategies in evaluating certain shared goals, we also observed significant differences in goals, even when participants evaluated the same visualization. Such disagreements were particularly common for infographic designs, such as [m], [n], or [q]. For example, consider visualization [m] in \autoref{fig:stimuli} in \autoref{appendix-stimuli}, an infographic titled ``NBA Lockout: A Play-by-Play of the Impact on the NBA,'' which visually explains the various stages and consequences of an NBA lockout. It uses a mix of cartoon depictions of players, team logos, and a timeline detailing the steps that occur during a lockout and how it affects games, players, and other stakeholders.

\begin{flushright}
\includegraphics[width=\linewidth]{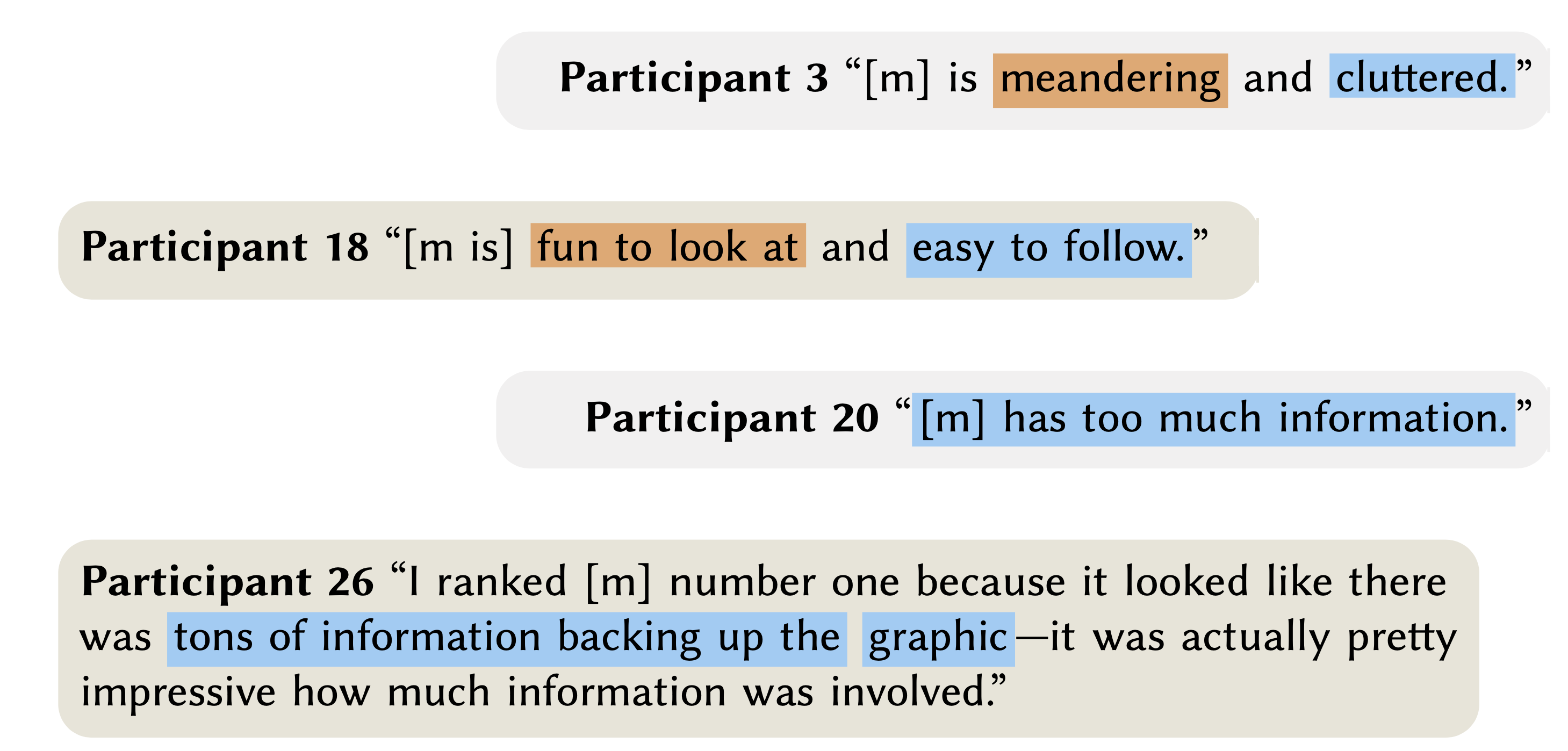}
\end{flushright}

The responses were notably polarized, reflecting diverse interpretations of its design and visual complexity, categorized as \design\ or \readability\ in our codebook, depending on the context. One participant, for instance, criticized [m] by stating, ``\textit{[m] is meandering and cluttered}'' [P3], suggesting that the layout was disorganized and difficult to follow. This view was echoed by others who found the presentation overwhelming or visually confusing.
In contrast, another participant had a completely different perspective, describing [m] as ``\textit{fun to look at and easy to follow}'' [P18]. For this individual, the same design elements that appeared cluttered to some were instead engaging and intuitive. This divergence highlights how subjective elements, such as visual appeal and ease of understanding, can vary widely across individuals, even within a single visualization.

In fact, P14 (who was generally opposed to infographic visualizations) specifically noted this conflict:



\begin{flushright}
\includegraphics[width=\linewidth]{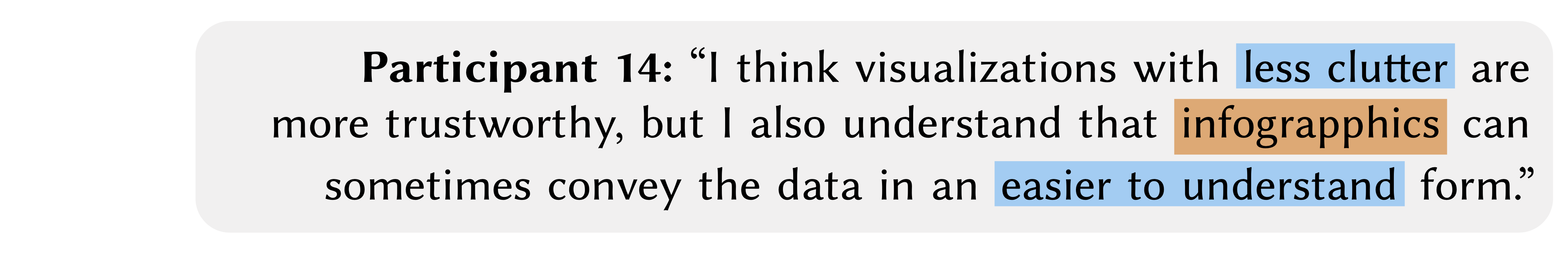}
\end{flushright}

This response not only summarizes the infographic discrepancy but also gives some insight into different perspectives on the \readability\ of visualizations. \ul{There appears to be a perceived trade-off between the potential accessibility of infographic visualizations and the perception of a lack of integrity behind the data represented.} This means that it is not only important to know what the audience wants in broad strokes (such as \readability), but it is also important to know if there are methods that in and of themselves are perceived to be less trustworthy. 
These contrasting responses illustrate the challenge of designing visualizations that cater to a wide audience. What one person perceives as cluttered, another may find informative. Thus, it is important to consider multiple perspectives in the design process and recognize that individual preferences and cognitive ability can significantly influence how visualization is interpreted~\cite{liu2020survey,ottley2022adaptive}.

\subsection{Theme 3: Holistic Review and Trends}
\label{sub:theme3}

Our third and final perspective is a top-down perspective on the trends that can be seen rather than individual comparisons of responses. When looking at the high-level patterns, it becomes clear that, despite the granular differences mentioned in \autoref{sub:theme2}, there are still important overall trends supported by our responses. 

\subsubsection{Participants were more likely to trust charts that they believed were clear and easy to understand} Although there is considerable variation in what individuals focus on when assessing the trustworthiness of a visualization, a few key factors consistently emerge. The most prominent factor mentioned was \clarity. This single keyword was mentioned by 31 of the 37 participants (83.8\%) in 91 of the 185 responses (49.2\%), making it the most frequently cited element by a large margin. Specifically, the \clarity\ keyword included responses that emphasized how easily participants could read or interpret the visualization.
\ul{The importance of clarity reflects a desire for visualizations to communicate information while minimizing confusion or cognitive load.}

\subsubsection{The type of chart impacted perceived trustworthiness.}
The second most frequently mentioned keyword was \vistype, cited by 18 of the 37 participants (48.6\%) in 42 of the 185 responses (22.2\%). 
This category referred to participants’ discussions about the specific format or style of the visualization, such as bar charts, line graphs, or more complex infographic-style visuals. 
\ul{This clearly implies that the type of visualization may play a role in shaping trust}, likely because certain formats may align better with participants’ preferences or expectations.

Interestingly, the most common discussions about \vistype~revolved around infographic-style visualizations, which often blend graphical elements, images, and text with data. These visualizations can be engaging and visually appealing, but participants expressed mixed opinions about their trustworthiness. While some viewed infographics as an effective way to summarize complex information, others were skeptical, feeling that the emphasis on design and aesthetics could detract from the presentation of accurate or comprehensive data. This lack of consensus on whether more ``data-centric'' visualizations are inherently more trustworthy than infographic-style ones underscores the diversity of user preferences and the subjective nature of trust. It also lends further evidence to the concept of External Inconsistency, discussed earlier in Section 2, where different individuals interpret the same visual elements in contrasting ways.

\subsubsection{People were more likely to trust when sources were cited and credible.}
The third most-used keyword was \source. This factor refers to explicitly citing the data or funding source behind a visualization. \textbf{Participants indicated that a cited source—particularly one they deemed reliable—significantly increased their trust in the data and the visualization.}
However, the relationship between source citation and trust is more nuanced than it might initially seem. Merely including a source does not automatically guarantee increased trust. As one participant pointed out in reference to visualization [n]:



\begin{flushright}
\includegraphics[width=\linewidth]{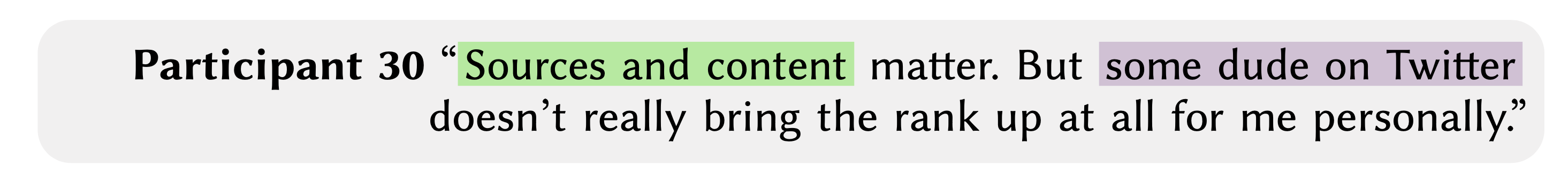}
\end{flushright}

This statement highlights the credibility of the source. In this case, the participant expressed skepticism toward the reliability of a Twitter user, suggesting that the perceived expertise or reputation of the source plays a critical role in determining trustworthiness. A source that is unfamiliar or perceived as biased can actually detract from trust rather than enhance it.
This perspective was echoed by other participants who emphasized their preference for ``reputable sources'' [P15] or ``reliable sources'' [P22] and underscores that trust in visualization is not solely based on the inclusion of a citation but is also influenced by participants' perceptions of the source's credibility.

\subsubsection{People tend to trust what they already know.}

The theme of \familiarity\ was also a common topic discussed by participants in their responses. This came in three kinds: topic, source, and unspecified, with the most common being topic. \topicfamiliarity\ in particular was used in 21 of the 185 responses (11.4\%) and by 12 of the 37 different participants (32.4\%). 
The usage was generally straightforward, explicitly calling upon prior knowledge held by the user. 

P23 heavily focused on prior knowledge and belief, as in the following example:



\begin{flushright}
\includegraphics[width=\linewidth]{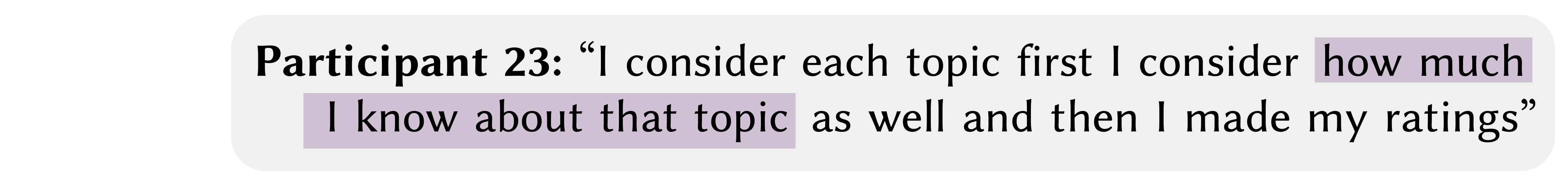}
\end{flushright}

\noindent
However, sometimes appeals to prior beliefs were implicit, as with P19:



\begin{flushright}
\includegraphics[width=\linewidth]{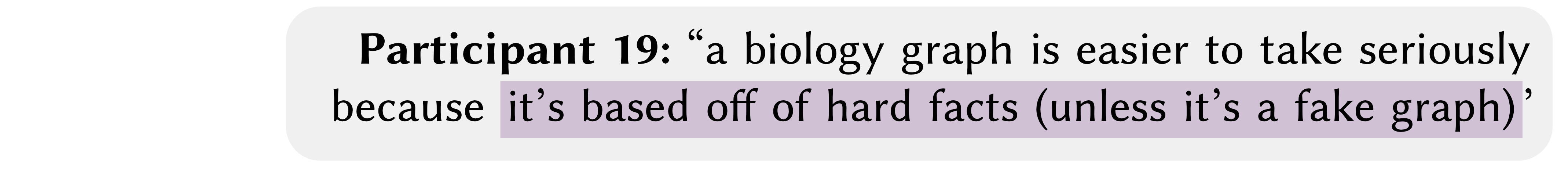}
\end{flushright}

These appeals are both drawing upon prior beliefs or biases held by the participants. In the case of P23, this is a more explicit appeal to their prior belief, whereas P19 may not be aware of the bias they are operating with (i.e., that scientific or technical topics are more ``based off of hard facts''). This is important, as it implies that \ul{the prior knowledge or assumptions about the topic may affect how a user will interact with and trust a visualization.}

\subsubsection{Aesthetics were important but often secondary.}

Finally, despite the importance of simplicity and clarity, there is also a contingent of 11 participants who explicitly mentioned the visual or aesthetic appeal of visualization in their trust perception. 
However, for many, this was a secondary concern. For example, P20 stated the following:



\begin{flushright}
\includegraphics[width=\linewidth]{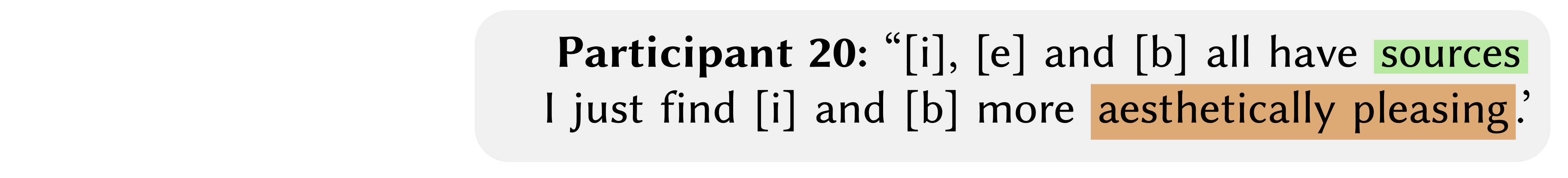}
\end{flushright}

For context, P20 ranked [i] and [b] as first and second rank, with [e] in third place, implying that P20 placed the most importance on sources but that their secondary concern was aesthetics. \ul{Aesthetics seem to be an important concern to our participants that can be superseded by other factors in the users' trust framework.}

\begin{figure*}
    \centering
    \includegraphics[width=.8\linewidth, alt={A paired bar chart showing the demographic differences in keyword categories by demographic variables (namely sex, age, education, and design experience).}]{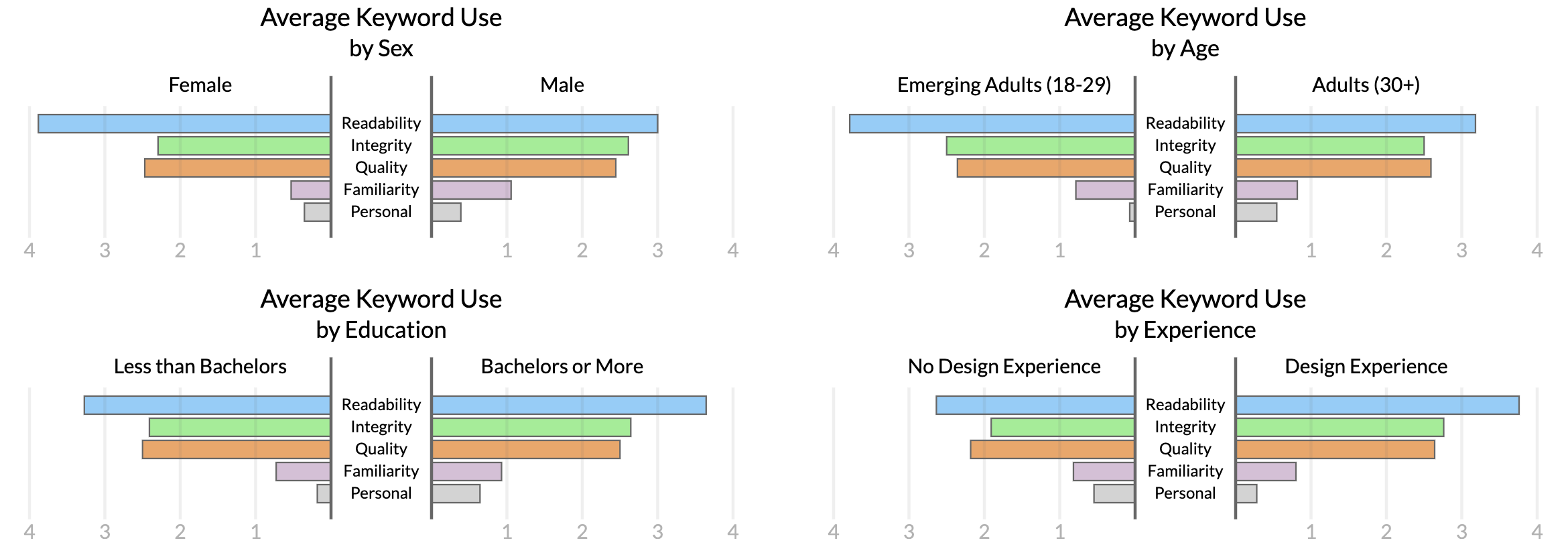}
    \caption{The frequency of different keyword categories broken down by education level, design experience, sex, and age. Each bar represents the average number of responses per participant that were associated with each category.}
    \label{fig:keywordfreq}
\end{figure*}

\subsection{Trust Rankings}
\label{sub:ranking}

Although the primary focus of this survey is on the textual responses elicited from users, it is still useful to take an exploratory look at the rankings provided for each set of visualizations to determine whether and how they align with the results above. This analysis was exploratory and targeted solely at finding patterns in the rankings that support or contradict the qualitative results discussed above. 

To that end, we performed a Rank Analysis using the sum-product of rankings for each visualization in each round. This is an application of weighted sums, using the rankings as weights \cite{churchman1954weighting, fishburn1967weighting}. As our design ranks from 1 (best) to 6 (worst), the visualization with the lowest sum-product is considered the highest ranked in a given round. \autoref{appendix-rankings} includes \autoref{fig:stimuli}, a visual of the distribution of rankings for each visualization by round. The rounds are also ordered from left to right by increasing sum-product.

The first result is that \ul{highly technical visualizations, such as [c] or [d], were consistently ranked very low}. This seems to align with our finding that \readability~and \familiarity~appear very important to many participants' trust perception. As these visualizations require a deep knowledge of their technical disciplines, it is likely that lacking this knowledge (and therefore being unable to parse the information in the chart) is at least partially responsible for these low ratings.

Conversely, \ul{simple charts like [b] and [i] consistently ranked very high}. In both rounds that contained these visualizations, these were the two highest-ranked visualizations. Similarly, other simple chart types, such as [g], [h], [a], and [e], were consistently ranked highly. The only time that one of these visualizations was not ranked in the top three was in Round 4, in which there were four of these visualizations ([b], [i], [g], and [e]) present. Round 5 was the only round where any of these charts (namely, [e]) was ranked lower than a less common chart type. This complements the interpretation that clarity and familiarity are important to trust perception and highlights that \vistype~appears to play a role.

Next, we looked at the infographic visualizations and found that \ul{infographics were generally ranked lower than the simple charts but higher than the technical charts}. Especially [m] and [n], the most recurring infographic visuals, were largely somewhere between these other groups. [k] also fell into this trend. This supports our finding that infographics can often be controversial. Some other infographics, such as [p] and [q], were ranked the lowest in their respective rounds. Still, these visualizations, in particular, were some of the most commonly referred to with regard to Intent. In other words, these visuals were sometimes seen as propagandistic or pushing an agenda, which highlights Intent as another seemingly important factor.

There were individual exceptions to these trends. For instance, visualization [o] ranked the highest in Round 5, despite being an infographic visualization, and even outranked [e]. Similarly, chart [f], despite being technical, was ranked third of six in Round 3, likely due to the competition of that round containing particularly unpopular visuals like [c] and [p].

Lastly, \ul{the rankings support our finding that participants seemed to have consistency in their evaluation process}. Chart [e] had the largest variance, being ranked between 2 and 4 on average in each round, though even this is relatively low considering that it was present in all five rounds. All other repeated visualizations consistently ranked within a one-ranking (like [c] or [d]) or two-ranking (like [m] or [n]) range. This implies that, despite changing contexts, participants often ranked visualizations consistently between rounds.

\subsection{Examining potential demographic predictors}
\label{sub:demo}

Prior work argues that demographic factors such as age, education, and design experience may influence how individuals perceive and process information~\cite{crouser2024building, elhamdadi2023vistrust}. For example, those with a higher education or more design experience might interpret visualizations differently than those with less exposure to data representation techniques~\cite{crouser2024building}. Similarly, older individuals may prefer simpler, more straightforward visualizations, while younger users may be more comfortable with complex or interactive visuals. 
Therefore, we examine these demographic factors to identify whether certain groups require different design considerations to build trust in visualizations. \autoref{fig:keywordfreq} summarizes the findings.

We found that education level, design experience, sex, and age did not significantly influence which design elements participants focused on when judging the trustworthiness of a visualization. This observation challenges the assumption that these demographic characteristics can shape individuals' trust perceptions of data visualizations. Instead, the evidence points to similar keyword usage across different groups, indicating that most people, regardless of background, prioritize the same design elements when evaluating the trustworthiness of visualizations, or at least that the inconsistencies between users are consistent across varying demographics.

We then ran $\chi^2$ tests for each keyword, removing those that did not meet the assumption of at least 5 instances per group. This resulted in 32 of the keyword-demographic pairs being tested, of which none were significant at the Bonferroni-corrected $\alpha$ level of 0.002. In other words, \ul{none of the analyses of individual keywords were able to reject the null hypothesis of no difference between demographic groups}.  

These results could imply that it may be possible to create design guidelines that resonate with a large segment of the general population, as suggested in \autoref{sub:theme3}. Since trust in visualizations appears predominantly based on consistent design elements like clarity, transparency, and visual encoding, designers may not need to significantly tailor their work for different demographic segments. This could streamline the design process, allowing for more focus on core principles of visualization design that ensure accessibility, simplicity, and neutrality rather than customizing visualizations to specific groups.

However, it's vital to recognize the limitations of this study, particularly with regard to the relatively small sample size of 37 participants. While the data suggests that demographic factors may not be strong indicators in this case, it is possible that larger or more diverse samples could reveal demographic influences that could not be detected in our sample size. Similarly, other demographic variables, such as socioeconomic conditions or visualization literacy, may also be responsible for the variation in response content and keyword use. Therefore, future research with a broader participant pool must confirm these findings or test other nuanced differences, such as visualization literacy~\cite{lee2016vlat,pandey2023mini} or personality factors~\cite{ottley2022adaptive,liu2020survey}. 

\section{Discussion}
\label{sec:discussion}

Along with the themes discussed in \autoref{sec:analysis} come some important takeaways from this work. These fall into two categories: guidelines for designers and takeaways for future research. 

\subsection{Designer Guidelines}
\label{sub:guidelines}

As discussed in \autoref{sec:intro}, a primary goal of this work is to create useful and specific guidelines for designers to follow to create trustworthy visualizations. Based on the results in \autoref{sec:analysis}, there are some clear takeaways that would be useful for designers to apply to their own trustworthy visualizations. \autoref{fig:teaser} contains the guidelines in a distilled, digestible form, and here we discuss the deeper explanations, rationales, and derivations behind each of the guidelines.

\subsubsection{Readability}

Looking through our responses, it is clear that our participants were most concerned with their ability to easily and effectively understand the data and message represented by the visualization. This aligns strongly with prior design guidelines~\cite{tufte20012e, kelleher2011guidelines, evergreen2013principles, midway2020principles} for effectiveness in data visualization communication, and we echo those guidelines here. 

\begin{quote}
    \textbf{Present Data Clearly.} Ensure that data and messages are easy to understand. Use simple, clean visualizations with clear labels to enhance trustworthiness.
\end{quote}

\subsubsection{Type of Visualization}

It is important to note, however, the disagreement between responses. The most polarizing topic in our survey was infographic visualizations. As discussed in \autoref{sub:theme2}, it appears that the context and audience influence the effectiveness of infographics. But our results also provide evidence that common chart types may be seen as more trustworthy, especially in comparison.
From this insight, along with other similar sentiments in the responses, we can generate an additional guideline for designers:

\begin{quote}
    \textbf{Choose the Right Type of Chart.} Bar and line graphs often convey a sense of transparency. Infographics can enhance readability, but some view them as an attempt to influence opinions.
\end{quote}

\subsubsection{Aesthetics}

Although participants rarely considered aesthetics to be the most important factor in their trust perceptions, the fact that over one-in-four participants (11 of the 37, 29.7\%) listed it as a factor still makes it a vital consideration. Designers should focus on making their visualizations clean and aesthetically pleasing.

\begin{quote}
    \textbf{Invest in Aesthetics.} Aesthetically pleasing designs increase trust. Invest time in polishing your chart and conveying professionalism but avoid overly cluttering your design.
\end{quote}

\subsubsection{Prior Knowledge}

As discussed in \autoref{sub:theme3}, people tended to trust what they knew. Therefore, prior biases or knowledge held by a viewer will impact their trust perception of a visualization. In our responses, this was most notable in the topic of the visualization, but it also extends to the data/design source and type of chart. In this way, we can see an important takeaway for designers:

\begin{quote}
    \textbf{Leverage Familiarity.} Tailor your design to what your audience knows. Use familiar topics, chart types, and formats to build trust.
\end{quote}

However, it is also important to note that a necessary corollary of ``people trust what they know'' is that the designer can always help them know more. For example, adding clear labels or putting a visualization in the context of titles, captions, or even broader articles or tutorials can take advantage of familiarity. We then posit that such practices will likely lead to higher trust when the topic or visual is novel and prior knowledge cannot be assumed. 

\begin{quote}
    \textbf{Educate Where Necessary.} When presenting novel or complex information, guide your audience with explanations, labels, or tutorials to foster understanding and trust. 
\end{quote}

\subsubsection{Source Citation}

Another clear takeaway discussed in \autoref{sub:theme3} is the explicit citation of sources, especially for the data in the visualization. Our participants were surprisingly sensitive to the presence or absence of source citations, and this was one simple aspect that many participants reported as leading to more trust. 

It is important to note, however, that not all sources are created equally. Although most participants who mentioned sources did so in a neutral way (i.e., only discussed the presence or absence of sources), several responses also mentioned prior opinions of the source. For instance, several participants mentioned ``who the sources were'' [P30], ``the quality and relevance of sources'' [P10], or specifically wanting ``reputable'' or ``official sources'' [P15]. This means there is more nuance than simply citing any source and expecting the same results.

\begin{quote}
    \textbf{Cite Credible Sources.} Always cite data sources to boost trust. Use trusted sources that resonate with your target audience for added confidence. 
\end{quote}

\subsection{Academic Takeaways}
\label{sub:academic}

As outlined in Section~\ref{sec:background}, much of the prior research on trust in data visualization has focused on defining trust or measuring it for specific visualization designs, often relying on subjective feedback, such as Likert scales. Our findings, in many ways, align with and validate this body of work.

For instance, Pandey et al.~\cite{pandey2023you} proposed five dimensions of visualization trust: Credibility, Clarity, Confidence, Reliability, and Familiarity. These dimensions are reflected in the themes that emerged from our qualitative analysis. \clarity~was the most frequently mentioned response category by a significant margin, and \familiarity~was also commonly discussed. Both \confidence~and \reliability~were explicitly referenced by multiple participants, while Credibility aligns closely with concepts like \dataintegrity~and \intent, both of which pertain to the data or designer’s trustworthiness. However, beyond the original framework, we uncovered nuanced emotional responses to different visualization types and specific aesthetic considerations. Additionally, we found that \familiarity~ \\ could be further dissected into distinct categories, such as \colorbox{table-purple}{Familiarity} \colorbox{table-purple}{with Topic}
versus \sourcefamiliarity. These granular distinctions highlight areas that are often overlooked in broader theoretical constructs.

Similarly, our findings resonate with the work of Elhamdadi et al.~\cite{elhamdadi2023vistrust}, which categorized trust into affective-based components (e.g., \visual, \aesthetic, \intuition) and cognition-based components (e.g., \clarity, \dataintegrity, \intent). This distinction was mirrored in our keyword analysis. Additionally, their work highlights the influence of individual characteristics on trust perceptions—a pattern corroborated by the variations we observed in participants' trust judgments. Moreover, aspects such as the perceived \objectivity~of the visualization and a general preference for \simple~designs suggest that users' trust evaluations may involve considerations beyond the binary of affective and cognitive components, indicating the need for more nuanced frameworks.

Our work also complements experimental research on trust in domains such as election integrity, public health, and misinformation~\cite{padilla2022multiple,yang2023midterms,yang2024dice}. While many of these studies focus on behavioral trust—such as whether users trust a specific chart or visualization—our study adopts a different approach. 
By asking participants to articulate their analytical trust processing (often related to as Type 2 processing frameworks~\cite{bancilhon2023eval, evans2013dual, kahneman_book}), we directly elicit their own perspectives on the feelings and thought processes underlying their trust judgments. 

This work does not seek to replace experimental studies, which provide valuable insights into trust behaviors and situational factors. Instead, it enriches this body of research by offering a complementary perspective, emphasizing the subjective, self-reported experiences of trust. By focusing on users’ articulated trust perceptions, we add depth to understanding how design elements and contextual factors influence trust in visualizations.

\subsection{Future Work}
\label{sub:future}

\paragraph{Incorporating Visualization Literacy Metrics}
This work's research implications will likely open many roads to future study. 
One such avenue for future work is to include visualization literacy in the analysis. Although this correlates with our survey's experience question, they are not the same metric. People with low visualization literacy can still have some experience making them, and people who have never made visualizations before could be skillful at reading charts. This literacy (or other metrics of knowledge around data or analysis) would likely be an important component of why different individuals' trust frameworks may vary. Future work would, therefore, do well to include this in their data (e.g., by including a Mini-VLAT test~\cite{pandey2023mini,lee2016vlat} during or around the survey).

\paragraph{Exploring Participant Consistency}
It is also important to note that the consistency that individual participants displayed in \autoref{sub:theme1} may be due to various factors. Firstly, it may result from a strong internal framework that is not dependent on the context of the visuals. However, it may also result from cognitive biases or especially memorable aspects of specific visual designs. This opens an opportunity for future work to examine how consistent these frameworks are, even when there are longer gaps between evaluations or visuals that are less unique or more numerous.  

\paragraph{Empirical Validation of Guidelines}
Other future works could include a more thorough literature review of trust findings to create guidelines similar to those presented here. This would be especially useful if there were additional empirical tests to show the effectiveness (or lack thereof) of the resulting guidelines. Such work would certainly help spread the accessibility of trust in data visualizations to designers who are outside of the data visualization research community. 
The guidelines that are provided in \autoref{fig:teaser} and \autoref{sub:guidelines} are based on the responses given by users in this survey. This evidence is meaningful, but it will be helpful to directly test these guidelines in a more empirical design to not only test self-reported trust perception but also test trust-related behaviors. Empirically testing these guidelines would give even more evidence to their validity. 

\subsection{Limitations}

The first and clearest limitation of this work is its exploratory nature. As we cast a broad net, many of the results should be further tested in future targeted work. For example, though we did not find any relationship between demographic variables and the factors identified by responses, there are several opportunities to further understand the potential role personal differences play. This could be done with a larger set of participants (for more statistical power) or by collecting more extensive demographic information, including visualization literacy, socioeconomic status, culture or national origin, and more.

In addition, although our methodology was decided on with the rationale discussed in \autoref{sub:survey}, there are other design choices with their own merits. Therefore, running experiments with different designs (e.g., interviews or forced-choice ranking) would be useful in further understanding trust. This survey also focused exclusively on static visualizations, specifically a small range of potential visualization types or designs. These results should be further validated by testing a similar design with a wider range and variety of data visualizations.

There is also the inherent limitation of using a survey in the first place. In this setting, we are looking at a controlled environment in which participants are explicitly asked to judge a visualization based on their trust perception in a quiz-like format. This context is drastically different from other contexts in which people are most likely to interact with visualizations (e.g., in news articles, journal submissions, social media, etc.).
Similarly, as this is an online survey conducted through Qualtrics with participants recruited through Prolific, there are inherent biases in sampling. Namely, participants will likely be those who spend more of their time online, which could inherently make those participants more familiar with data visualizations. This research also focused on US-based English speakers, and future work would do well to expand this work further to other languages or locales.

In addition, our keyword coding was completed by two authors and judgmentally assigned to responses. This means that, given another encoder, this encoding could turn out differently. While the ability to make (and more importantly, justify) these judgments is a benefit of using manual, human coding, it is also important to note that different coders (including automated algorithms) may come to different judgments about how to apply the codes to responses. 

Finally, another difficulty with the survey setup is that it does not allow further information or follow-up questioning. If there was an unclear statement that required the aforementioned judgment, there was no way to clarify with the participant what exactly they intended. An interview style could alleviate these concerns, though it would also be more time- and resource-intensive.

\section{Conclusion}
\label{sec:conclusion}

In this work, we build upon prior research in data visualization trust by gathering a new perspective directly from users on how they perceive trust. This user-centric approach uncovered several key insights into trust perception in the context of data visualizations.
One significant finding is that participants appear to use consistent internal frameworks to assess the trustworthiness of visualizations. The same evaluation criteria were repeatedly applied across different visualizations, and similar assessments were made even when the context or timing changed.
However, we also found no universal formula for designing a trustworthy visualization. Participants reported diverse frameworks for evaluating trust, with variation in both the factors considered and the strategies used for assessment. Trust in visualization is a highly personal decision, with rationales differing from one individual to another. This underscores the importance of designers understanding their target audience, as opinions on what constitutes a trustworthy visualization can vary widely.
Despite this variability in trust perception, several common trends emerged from the study, such as the critical role of clear, readable designs, disclosing source citations, thoughtful aesthetic choices, and using familiar topics and chart types.
We translated these insights into specific guidelines for designers aiming to create trustworthy visualizations. \ul{We recommend that designers: Present Data Clearly, Choose the Right Type of Chart, Invest in Aesthetics, Leverage familiarity, Educate Where Necessary, and Cite Credible Sources.}
By following these guidelines, designers can have greater confidence in creating visualizations that are perceived as trustworthy.

\begin{acks}

This work was partially supported by the National Science Foundation (NSF) under Award No. 2142977 and 2330245, which funds the Engineering Research Center for Carbon Utilization Redesign through Biomanufacturing-Empowered Decarbonization (CURB).

\end{acks}

\balance

\bibliographystyle{ACM-Reference-Format}
\bibliography{FINAL_SUBMISSION}

\clearpage
\onecolumn

\appendix 

\section{Stimuli}
\label{appendix-stimuli}

\begin{figure*}[h]
\centering
\includegraphics[width=.9\linewidth, alt={A compilation of the 17 visualizations used in this study, with letter labels for ease of reference throughout the paper.}]{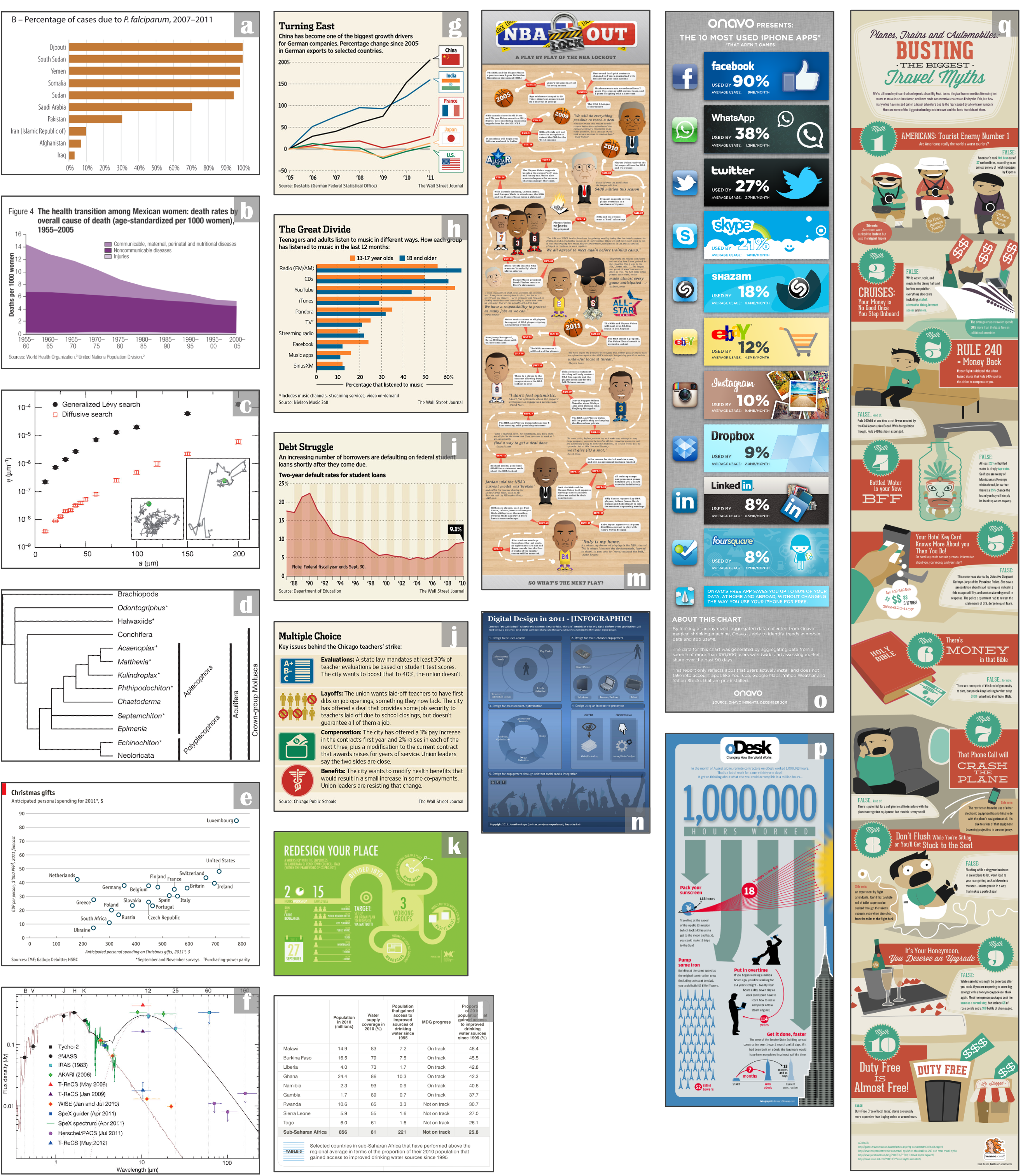} 
\caption{
The 17 visualizations used in the study\cite{pandey2023you,borkin2013makes}. These visualizations include line, bar, infographic, and more visualizations from news networks, scientific publications, and government sources. Each is given a letter ID from [a] through [q]. 
}
\label{fig:stimuli}
\end{figure*}

\clearpage

\section{Code Book Definitions}
\label{appendix_codebook}

\begin{table*}[h]
    \centering
    \caption{Codebook Definitions with Number of Participants and Responses}
    \begin{tabular}{|l|l|l|l|}
        \hline
        \textbf{Keyword} & \textbf{Part. (37)} & \textbf{Resp. (185)} & \textbf{Definition} \\
        \hline
        \hblue{Clarity} & 31 (83.8\%) & 91 (49.2\%) & The response is concerned with how clear or understandable the \\
        &&&visualization or message is. \\
        \hline
        \hblue{Simple} & 15 (40.5\%) & 29 (15.7\%) & The response explicitly prefers or trusts simpler visualizations. \\
        \hline
        \hblue{Complex} & 4 (10.8\%) & 5 (2.7\%) & The response explicitly prefers or trusts more complex visualizations. \\
        \hline
        \hgreen{Data Integrity} & 15 (40.5\%) & 21 (11.4\%) & The response is concerned with the perceived integrity of the data \\
        &&&behind the visualization. \\
        \hline
        \hgreen{Intent} & 12 (32.4\%) & 17 (9.2\%) & The response is concerned with malicious intent by the designer. \\
        \hline
        \hgreen{Source} & 11 (29.7\%) & 36 (19.5\%) & The response explicitly mentions the presence or lack of source citations. \\
        \hline
        \hgreen{Labeling} & 3 (8.1\%) & 7 (3.8\%) & The response specifically mentions (positively or negatively) the data \\
        & & & labels present or absent on the visualization. \\
        \hline
        \hgreen{Objectivity} & 2 (5.4\%) & 8 (4.3\%) & The response explicitly refers to the objectivity (or subjectivity) of a  \\
        & & & visualization or designer. \\
        \hline
        \horange{Vis Type} & 18 (48.6\%) & 41 (22.2\%) & The response explicitly refers to the kind of visualization used (e.g., \\
        & & & bar chart, line chart, infographic, etc.) \\
        \hline
        \horange{Visual} & 16 (43.2\%) & 25 (13.5\%) & The response specifically mentions (positively or negatively) any visual  \\
        \horange{Elements} & & & design elements present or absent on the visualization (excluding those \\
        & & & otherwise coded) or the general design without specification. \\
        \hline
        \horange{Aesthetics} & 11 (29.7\%) & 13 (7.0\%) & The response is concerned with the aesthetic appeal of the visualization.  \\
        \hline
        \horange{Professional} & 5 (13.5\%) & 8 (4.3\%) & The response explicitly uses the term ``Professional'' (or a close synonym, \\
        & & & such as ``official'') in a positive context with regard to trustworthiness. \\
        \hline
        \horange{Academic} & 3 (8.1\%) & 3 (1.6\%) & The response explicitly uses the term ``Academic'' (or a close synonym, \\
        & & & such as ``scientific'') in a positive context with regard to the participant's \\
        &&& trust perception. \\
        \hline
        \hpurple{Familiarity} & 12 (32.4\%) & 21 (11.4\%) & The response either explicitly refers to or implicitly uses prior knowledge \\
        \hpurple{with Topic} & & & of the topic the visualization represents. \\
        \hline
        \hpurple{Familiarity} & 4 (10.8\%) & 5 (2.7\%) & The response either explicitly refers to or implicitly uses prior knowledge \\
        \hpurple{with Source} & & & of the data source or designer. \\
        \hline
        \hpurple{Unspecified} & 1 (2.7\%) & 2 (1.1\%) & The response explicitly refers to prior knowledge but is unclear about the \\
        \hpurple{Familiarity} & & & subject of that prior knowledge. \\
        \hline
        \hgray{Confidence} & 5 (13.5\%) & 5 (2.7\%) & The response explicitly states that the confidence the participant feels is a \\
        & & & factor in the participant's trust perception. \\
        \hline
        \hgray{Reliability} & 4 (10.8\%) & 5 (2.7\%) & The response explicitly states that the perceived reliability of a visualization \\
        & & & is a factor in the participant's trust perception. \\
        \hline
        \hgray{Intuition} & 1 (2.7\%) & 2 (1.1\%) & The response explicitly states that intuition or ``gut feeling'' is a factor in \\
        & & & the participant's trust perception. \\
        \hline
        \hgray{Relevance} & 1 (2.7\%) & 1 (0.5\%) & The response explicitly states that the visualization's relevance to the \\
        & & & participant is a factor in the participant's trust perception.  \\
        \hline
        \hgray{Importance} & 0 (0.0\%) & 0 (0.0\%) & The response explicitly states that the perceived importance of a \\
        & & & visualization is a factor in the participant's trust perception. Only used by \\
        &&& P6, whose responses were removed.\\
        \hline
        Trust & 1 (2.7\%) & 1 (0.5\%) & The response is limited to an appeal to the word ``trust'' (e.g., ``Order I trust \\
        & & & them.'' [P6]) without elaboration. \\
        \hline
        No Keywords & 3 (8.1\%) & 3 (1.6\%) & The response does not provide any new information (e.g., ``arranging them \\
        & & & in order'' [P27] or ``Same method as before'' [P8]). \\
        \hline
    \end{tabular}
    \label{tab:codedefs}
\end{table*}

\clearpage

\section{Visualization Rankings}
\label{appendix-rankings}

\begin{figure*}[h!]
\centering
\includegraphics[width=.9\linewidth, alt={A series of histograms displaying the ranking distribution of each visualization in each round of the survey.}]{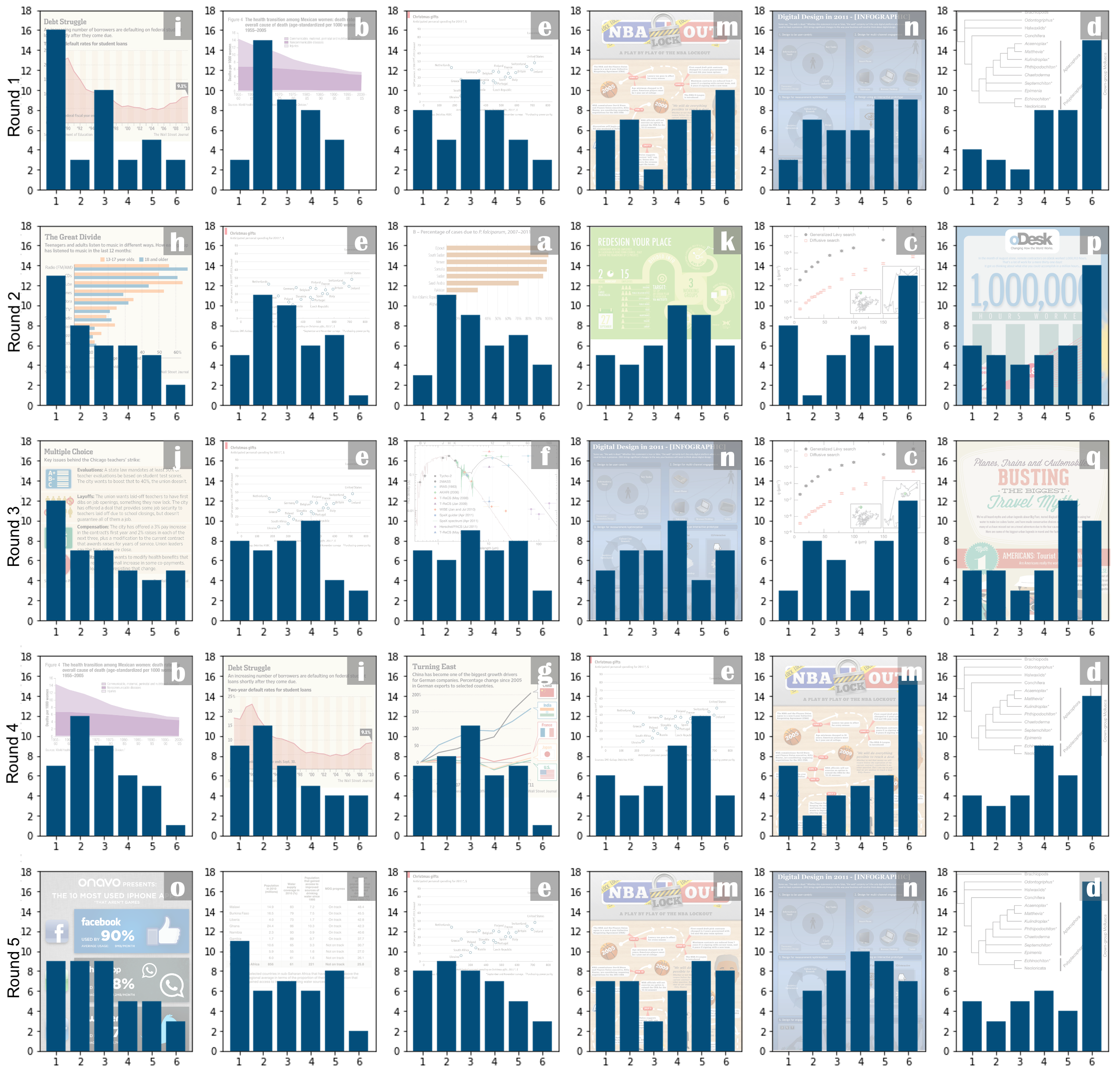} 
\caption{
These histograms represent the rankings given to each visualization in each round (with 1 as most trusted and 6 as least trusted). The visualizations in each round are ordered by sum-product score, with the left having the best (lowest) score and right having the worst (lowest) score.
}
\label{fig:ranking}
\end{figure*}

\end{document}